%% file: main.tex
\documentclass[final,1p]{elsarticle}

\biboptions{sort&compress,square}
\usepackage{empheq}
\usepackage{todonotes}
\usepackage{enumitem}
\usepackage[running]{lineno}

\def\tdot   {\;\cdot\!\!:}

\input{packages_style_defs.tex}

\newdefinition{rmk}{Remark}

\journal{arXiv}

\title{Dissipation-consistent modelling and classification of extended plasticity formulations}
\author[label1,label2,cor]{A.T. McBride}
\author[label2]{B.D. Reddy}
\author[label1,label3]{P. Steinmann}
 \cortext[cor]{Corresponding author.}

 \address[label1]{Glasgow Computational Engineering Centre, University of Glasgow, United Kingdom}
 \address[label2]{Centre for Research in Computational and Applied Mechanics, University of Cape Town, South Africa}
  \address[label3]{Chair of Applied Mechanics, Friedrich-Alexander University of Erlangen--Nuremberg, Germany}

\begin{document}
\begin{frontmatter}

\begin{abstract}
A unified classification framework for   models of extended plasticity is presented. 
The  models include well known micromorphic and strain gradient plasticity formulations.
A unified treatment is possible due to the representation of strain gradient plasticity as an Eringen-type micromorphic continua.
The classification is based on the form of the energetic and dissipative model structures and exploits the framework of dissipation-consistent modelling to elucidate the flow relation and yield condition. 
Models are identified as either serial or parallel. 
This designation is also applicable to familiar  models of classical plasticity. 
Particular attention is paid to the rate-dependent problem arising from the choice of a smooth dissipation potential.
The inability to locally determine the region of admissible stresses for the non-smooth (rate-independent) parallel models of plasticity is made clear.
\end{abstract}

\end{frontmatter}

\section{Introduction}

 Classical theories of plasticity are unable to account for the experimentally-observed, size-dependent response exhibited by structures at the mesoscopic scale.
 Starting with the early works of \citet{Aifantis1984, Aifantis1987}, extended models of plasticity have been actively developed and analysed over the past three decades to remedy these and related deficiencies associated with the pathological localisation of softening problems.
 A significant proportion of extended plasticity models are members of either the gradient or micromorphic frameworks, or indeed both.
 Micromorphic continua are  characterised by additional degrees of freedom at each continuum point \citep{Mindlin1964, Toupin1964, Eringen1999}.
 By contrast, gradient continua possess higher gradients of their primary fields.
 Important examples of micromorphic and gradient plasticity theories include \citep{Steinmann1991, deBorst1991, Grammenoudis2001} and \citep{Fleck1997, Fleck2001, Gurtin2002, Gudmundson2004, Gurtin2005, Gurtin2005b, Nix1998}, respectively.
For  classifications of extended models of inelasticity, the reader is referred to \citep{Kirchner2005, Hirschberger2009, Forest2009}.

Despite the considerable work on models of extended plasticity, open questions and challenges remain.
For example, our recent work \citep{Carstensen2017} on a small-strain theory of strain gradient plasticity due to \citet{Gurtin2005}, detailed the implications of the structure of the energetic and dissipative components of the model on the nature of yield and subsequent plastic flow.
An objective of the current work is to present a novel classification scheme based on the energetic and dissipative structures for models of extended plasticity.
The classification scheme also clarifies important  features of various models of plasticity, for example, the local or global nature of the region of admissible stresses. 
Furthermore, we significantly extend our previous work \citep{Carstensen2017} to include the important rate-dependent case obtained when the dissipation potential is chosen to be smooth.

Models are classified here as being examples of either \emph{serial plasticity} or \emph{parallel plasticity}.
This novel distinction is based upon the choice of the energy storage potential and the dissipation potential, and has important consequences for the determinability of the (macroscopic) stress and hence the region of admissible stresses.
We show that the essence of this distinction can be explained in the familiar setting of local plasticity (i.e.\ in the absence of strain-gradient terms).
Models of \emph{local serial plasticity} and \emph{local parallel plasticity} are constructed from simple rheological units and analysed for both smooth and non-smooth dissipation potentials, corresponding to the rate-dependent and rate-independent problem, respectively. 
The key features of the local models are shown to hold for the  models of extended plasticity. 
Importantly, we show that the stress is fully determined from the elastic law in the local serial model.
This is not however the case for the local parallel plasticity model.

The framework of generalised dissipative materials \citep{Biot1965, Ziegler1963, Halphen1975} is exploited in this work to provide the basic procedure to determine the structure of the boundary value problem and the internal variable evolution relations \citep[see e.g.][]{Miehe2011, Miehe2014a, Miehe2014b}.
The framework of generalised dissipative materials is related to the primal formulation of plasticity \citep{Han1997, Carstensen1999, Carstensen2000, Han2010} central to our previous contribution \citep{Carstensen2017}.
Primal formulations of strain gradient plasticity have received considerable attention recently and their variational structure analysed \citep[see e.g.][]{Djoko2007, Reddy2008, Reddy2011_partI, Reddy2011_partII, Reddy2012erratum}.

The proposed classification scheme is applicable to  models of extended plasticity. 
A unified scheme is achieved by considering the chosen extended models as variants of an Eringen-type micromorphic continuum. 
This allows for the classification of a range of important micromorphic and strain gradient plasticity models. 
The extended models are categorised as either \emph{micromorphic serial plasticity} or \emph{micromorphic parallel plasticity}.
The proposed classification scheme provides considerably more structure than the designations \emph{energetic} or \emph{dissipative} (or variants thereof) used in many strain gradient plasticity models \citep[see e.g.][]{Fleck2009_partI, Fleck2009_partII, Carstensen2017}.

The classification  of micromorphic parallel plasticity encapsulates the important \citet{Gurtin2005} model and extensions thereof.
In order to elucidate the implications of the choice of the energetic and dissipative structures for the  \citeauthor{Gurtin2005}  model, we introduce three categories of micromorphic parallel plasticity: (i) the combined energetic-dissipative case, (ii) the hybrid energetic-dissipative case, and (iii) the fully-dissipative case.
The fully-dissipative, rate-independent  case was the focus of our previous work \citep{Carstensen2017}.
That work was motivated, in turn, by  the recent work of \citet{Fleck2014} on the behaviour of strain-gradient models under conditions of non-proportional loading.
Of particular interest in that work is the appearance of a so-called ``elastic gap'' in the response at a material point undergoing plastic flow due to the instantaneous halting of plastic flow on part of the boundary, or due to non-proportional loading.
We clearly show here that a local treatment of the \emph{fully-dissipative problem} is possible provided the dissipation potential is chosen to be smooth.
This is a new and important result and supports the numerical findings presented in \citep{Fleck2014}.
The analysis of the fully-dissipative case with a non-smooth dissipation potential differs from that in \citep{Carstensen2017} as we do not resort to a spatial discretisation from the onset.

The structure of the presentation is as follows.
 The general setting of dissipation-consistent modelling based upon standard dissipative materials  is presented in \sect{sec_general_setting}.
 This abstract presentation is then applied to the familiar problem of local plasticity in \sect{sec_local_plasticity}.
 Here models of local serial  plasticity and local parallel  plasticity are introduced.
 \sect{sec_micromorphic_gradient_plasticty} introduces micromorphic continua as a framework for strain gradient plasticity.
 The models of micromorphic serial plasticity and micromorphic parallel plasticity are presented in \sect{sec_micro_serial} and \sect{sec_micro_parallel}, respectively.
 Conclusions are drawn and extensions proposed in \sect{sec_conclusion}.

\section*{Notation and results}

Consider a continuum body occupying a domain $\Omega$ with boundary $\partial \Omega$.
The outward unit normal to $\partial \Omega$ is denoted by $\b{n}$.
A continuum point at position $\b{x} \in \Omega$ undergoes a displacement at time $t$ denoted by $\b{u}(\b{x},t)$.
The deformations are assumed infinitesimal.

The respective scalar products of arbitrary vectors ($\b{a}$ and $\b{b}$), second-order tensors ($\b{\sigma}$ and $\b{\varepsilon}$) and  third-order tensors ($\b{\eta}$ and $\b{\gamma}$)  are defined by
\begin{align*}
  \b{a} \cdot \b{b} = a_i b_i \, ,
  &&
  \b{\sigma} : \b{\varepsilon} = \sigma_{ij} \varepsilon_{ij} \, ,
  &&
  \b{\eta} \tdot \b{\gamma} = \eta_{ijk} \gamma_{ijk} \, ,
\end{align*}
where summation over repeated indices is implied.
A Cartesian coordinate system is used throughout.

The generalised scalar product between two arbitrary generalised measures $\b{A}$ and $\b{B}$ is denoted by $\b{A} \circ \b{B}$.
For example, if $\b{A} := \{\b{\sigma}, \b{\eta} \}$ and $\b{B} := \{\b{\varepsilon}, \b{\gamma} \}$, then
\begin{align*}
  \b{A} \circ \b{B} := \sigma_{ij} \varepsilon_{ij} + \eta_{ijk} \gamma_{ijk} \, .
\end{align*}
Due to its prominent role in the theoretical developments that follow, the scalar product of a driving force and the conjugated set of internal variables is distinguished from the generalised scalar product, denoted by an open circle, by using a closed circle, i.e.\ $\bullet$.

Differentiation of an arbitrary function $f(\b{A})$ with respect to a variable $\b{A}$ is denoted by $f_{,\b{A}}$.
Similarly, the subdifferential (defined in \eqn{subdifferential}) of an arbitrary function $f(\b{A})$  with respect to a variable $\b{A}$ is denoted by $f_{;\b{A}}$.

The spatial gradients of arbitrary scalar, vector, and tensor fields are respectively defined by
\begin{align*}
  \nabla a = \dfracp{a}{\b{x}} \, , &&
  \nabla \b{a} = \dfracp{\b{a}}{\b{x}} \, , &&
  \nabla \b{A} = \dfracp{\b{A}}{\b{x}} \, ,
\end{align*}
where $[\nabla \b{A}]_{ijk} = \partial A_{ij} / \partial x_k$.
The symmetric gradient operator is defined by $\nabla^\sym \b{a} = [1/2][\nabla \b{a} + [\nabla  \b{a}]\trns]$.

Standard results from convex analysis required for the presentation are provided in \ref{sec_convex_analysis}.

\section{The general setting of dissipation-consistent modelling} \label{sec_general_setting}

The structure of the constitutive equations and the inelastic evolution relations for a broad range of inelastic physical processes can be obtained from the general setting of dissipation-consistent modelling based on the notion of standard dissipative materials \citep[see e.g.][]{Ziegler1963, Biot1965, Germain1973, Halphen1975}.
The general setting presented in this section provides the basis for the investigation of both local and micromorphic plasticity that follows.

\subsection{The dissipation inequality}

We denote by $\b{S}$ and $\b{E}$  generalised stress (kinetic) and strain (kinematic) measures, respectively.
The set of internal (hidden) variables quantifying inelastic processes  is denoted by $\b{I}$.
The generalised stress $\b{S}$ is further additively decomposed into energetic  and dissipative parts as follows
\begin{align*}
  \b{S} = \b{S'}  + \b{S''} \, . 
\end{align*}
Note that $\b{A'}$ and $\b{A''}$ denote respectively the energetic and dissipative parts of an arbitrary variable $\b{A}$.

Within the isothermal setting assumed here, the dissipation density $d$ per unit  volume of $\Omega$ is defined by
\begin{align}
  d:= \b{S} \circ \dot{\b{E}} - \dot{\psi} \geq 0  \, , \label{d_1}
\end{align}
where $\psi$ is the energy storage potential (i.e.\ the Helmholtz energy density)  parametrised by $\b{E}$ and $\b{I}$, that is
\begin{align}
  \psi = \psi(\b{E},\b{I}) \, .
  \label{psi}
\end{align}
For simplicity, but without loss of generality, we assume the energy storage potential to be henceforth a quadratic function in the generalised strain $\b{E}$.

\subsection{The energy storage potential}

Substitution of expression \eqref{psi} for the energy storage potential $\psi$  into the dissipation inequality \eqref{d_1}, and employing the standard Coleman--Noll procedure \citep{Coleman1963}, yields the relation for the energetic stress as
\begin{align}
  \b{S'} := \psi_{,\b{E}} = \b{S} - \b{S''}\, . \label{S'_from_psi}
\end{align}
Thus $\psi$ serves as the potential for the energetic stress $\b{S'}$.

The \emph{reduced dissipation inequality} thus follows from the dissipation inequality \eqref{d_1} as
\begin{align}
  d = \b{S''} \circ \dot{\b{E}} - \psi_{,\b{I}} \bullet \dot{\b{I}}   \geq 0 \, .
  \label{d_2}
\end{align}
From the reduced dissipation inequality, it is apparent that the energy storage potential also serves as the potential function for the energetic  driving force $\b{X'}$ defined by
\begin{align}
  \b{X'} := \psi_{,\b{I}}
  && \text{where we define} &&
  \b{X''} :
  = -\b{X'} \, . \label{X'_and_X''}
\end{align}
The dissipative driving force $\b{X''}$ follows from Biot's relation \citep{Halphen1975} in \subeqn{X'_and_X''}{b} as the negative of the energetic driving force $\b{X'}$.

Finally, the reduced dissipation inequality \eqref{d_2} can be expressed as
\begin{align}
  d = \b{S''} \circ \dot{\b{E}} + \b{X''} \bullet \dot{\b{I}}   \geq 0 \, . \label{d_3}
\end{align}
In this format, the reduced dissipation inequality  identifies the dissipation conjugate pairings  as $ \b{S''} \leftrightarrow \dot{\b{E}}$ and $\b{X''} \leftrightarrow \dot{\b{I}}$.

\subsection{The dissipation potential}

The dissipation potential is denoted by $\pi$ and parametrised by $\dot{\b{E}}$ and $\dot{\b{I}}$, that is
\begin{align*}
  \pi = \pi(\dot{\b{E}}, \dot{\b{I}}) \, .
\end{align*}
The dissipation potential provides the required structure for the dissipative quantities $\{ \b{S''}, \b{X''} \}$ in the reduced dissipation inequality \eqref{d_3}.
The dissipation potential is a \emph{gauge} and hence satisfies the three properties listed in \eqn{gauge}.
Endowed with these properties, the dissipation potential characterises a \emph{standard dissipative material}.
A schematic of a non-smooth dissipation potential, for the restricted case where $\pi = \pi(\dot{\b{E}})$, is shown in \fig{fig_pi_pistar}(a).

For a standard dissipative material, the dissipative stress $\b{S''}$ and the dissipative driving force $\b{X''}$ are in the subdifferential of the dissipation potential (see \fig{fig_pi_pistar}(c)).
That is,
\begin{align}
  \b{S''} \in \pi_{;\dot{\b{E}}}
  && \text{and} &&
  \b{X''} \in \pi_{;\dot{\b{I}}} \, .
  \label{diss_S_X_in_pi}
\end{align}


\subsection{The dual dissipation potential}

The dual dissipation potential $\pi^\star$, the convex conjugate to $\pi$, is given by the Legendre--Fenchel transformation (defined in \eqn{legendre_fenchel}) as
\begin{align}
  \pi^\star(\b{S''}, \b{X''}) =
  \sup_{\dot{\b{E}},\dot{\b{I}}}
  \{
  \b{S''} \circ \dot{\b{E}} + \b{X''} \bullet \dot{\b{I}}
   -  \pi( \dot{\b{E}}, \dot{\b{I}}) \} \, . \label{legendre_fenchel_2}
\end{align}
The dual dissipation potential serves as the potential for the rate of the generalised strain $\dot{\b{E}}$  and the rate of the internal variables $\dot{\b{I}}$.
That is,
\begin{align*}
  \dot{\b{E}}  \in \pi^\star_{;\b{S''}}
  && \text{and} &&
  \dot{\b{I}} \in \pi^\star_{;\b{X''}} \, .
\end{align*}

\begin{rmk}
  If the dissipation potential $\pi$ satisfies the requirements of a gauge, and the material can therefore be classified as standard dissipative, then the reduced dissipation inequality \eqref{d_3} is automatically satisfied.
  To see this, note that positively homogeneous functions (defined in \eqn{positive_homogeneity}) are characterised by  Euler's Theorem for homogeneous functions.
  This  implies that
  \begin{align*}
    \b{S''} \circ \dot{\b{E}} + \b{X''} \bullet \dot{\b{I}} = k \pi(\dot{\b{E}}, \dot{\b{I})} \, ,
  \end{align*}
  for $k \geq 1$.
  The convexity of $\pi$, property \eqref{gauge_i}, and property \eqref{gauge_ii} imply that
  \begin{gather*}
    0 \geq \pi(\dot{\b{E}}, \dot{\b{I})} - \b{S''}\circ \dot{\b{E}}  - \b{X''}\bullet \dot{\b{I}} \, , \\
    \implies
    d = \b{S''}\circ \dot{\b{E}}  + \b{X''}\bullet \dot{\b{I}}
    = k \pi(\dot{\b{E}}, \dot{\b{I})}
     \geq \pi(\dot{\b{E}}, \dot{\b{I})} \geq 0 \, .
  \end{gather*}
  
  For the choice of a non-smooth dissipation potential (the rate-independent problem) $k \equiv 1$ and the dissipation density $d$ and the dissipation potential $\pi$ coincide. 
  This is not the case for a smooth dissipation potential  (the rate-dependent problem) where, in general, $d \not \equiv \pi$.
  \qed
\end{rmk}

\begin{figure}[h]
  \centering
  \includegraphics[width=\textwidth]{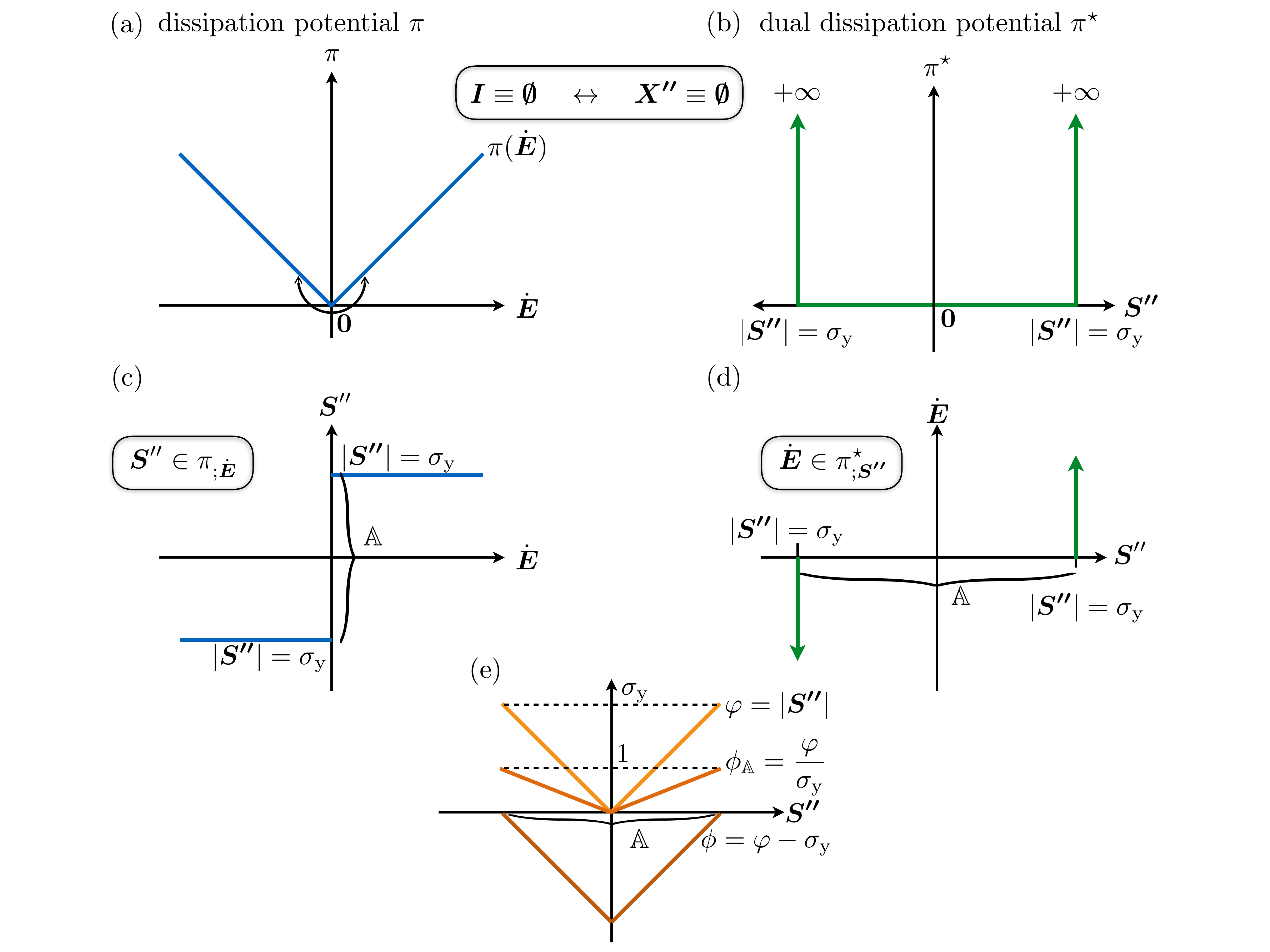}
  \caption{A schematic of the relation between the non-smooth dissipation potential $\pi$ in (a) and the corresponding dual dissipation potential $\pi^\star$ in (b), for the restricted case where $\pi = \pi(\dot{\b{E}})$ and $\pi^\star = \pi^\star(\b{S''})$.
  The relation for the dissipative stress $\b{S''}$ in terms of the dissipation potential $\pi$, and the generalised strain rate $\dot{\b{E}}$ in terms of the dual dissipation potential $\pi^\star$ are shown in (c) and (d), respectively.
  For this restricted case, the  region of admissible stresses $\mathbb{A} = \mathbb{A}(\b{S''})$ has as its limit the yield stress $\sigma_\text{y}$.
  The interior of the region of admissible stresses is denoted by $\interior{\mathbb{A}}$.
  The equivalent stress function $\varphi$, the yield function $\phi$ and the canonical yield function $\phi_\mathbb{A}$ are depicted in (e).}
  \label{fig_pi_pistar}
\end{figure}

\begin{rmk}
As mentioned, the rate-independent formulation corresponds to the choice of a non-smooth dissipation potential.
The rate-independent problem is important as it represents the limit of the rate-dependent theory. 
Some key results concerning the \emph{rate-independent} problem are now summarised. 

The admissible region $\mathbb{A} = \mathbb{A}(\b{S''},\b{X''})$ is a closed convex set.
The boundary of $\mathbb{A}$, denoted by $\bdy{\mathbb{A}}$, is the level set $\sigma_\text{y}$ of the convex function $\varphi(\b{S''},\b{X''}) = \sigma_\text{y}$, where $\sigma_\text{y}>0$, and $\varphi$ is termed the \emph{equivalent stress function}.
The equivalent stress function $\varphi$ can be expressed as a gauge on $\mathbb{A}$, and in this form is distinguished by the notation  $\phi_\mathbb{A}$ and termed the \emph{canonical yield function} (see \ref{sec_polar_functions}).
The yield function, denoted by $\phi$, is defined by
\begin{align*}
  \phi := \varphi - \sigma_\text{y} \leq 0 \, ,
  \intertext{and the canonical yield function by} 
  \phi_\mathbb{A} := \dfrac{\varphi}{\sigma_\text{y}} \leq 1 \, .
\end{align*}
For a schematic depiction of the various yield functions and the equivalent stress function, for the restricted case where $\pi = \pi(\dot{\b{E}})$, see \fig{fig_pi_pistar}(e).

For a non-smooth dissipation potential $\pi$, the dual dissipation potential $\pi^\star$ is the indicator function (see the definition in \eqn{indicator_function}) of the  admissible region $\mathbb{A}$ of generalised stresses and driving forces, that is,
\begin{align*}
  \pi^\star(\b{S''},\b{X''})
  = I_\mathbb{A}(\b{S''},\b{X''})
  = \begin{cases}
  0 & \text{if } \{\b{S''},\b{X''}\} \in \mathbb{A} \\
  +\infty & \text{otherwise}
  \end{cases} \, .
\end{align*}

The support function of the region $\mathbb{A}$, denoted by $\sigma_\mathbb{A}$ and defined in \eqn{support_function}, is given by
\begin{align*}
  \sigma_\mathbb{A}(\dot{\b{E}},\dot{\b{I}}) = \sup_{ \b{S''},\b{X''}}
  \{
  \b{S''} \circ \dot{\b{E}} + \b{X''} \bullet \dot{\b{I}}
  \} \, , \quad \{ \b{S''},\b{X''}\} \in \mathbb{A} \, .
\end{align*}
The dissipation potential $\pi$ is therefore the support function of the region $\mathbb{A}$.

Thus, from \eqn{indicator_support_relation}, the support function is conjugate to the indicator function, that is
\begin{align*}
I^\star_\mathbb{A} = \pi^{\star\star} = \sigma_\mathbb{A} = \pi
 \, .
\end{align*}

Then, from \eqn{polar_functions}, for $\{ \b{S''},\b{X''} \} \in \mathbb{A}$ and $\b{S''} \in \pi_{;\dot{\b{E}}}$ and $\b{X''} \in \pi_{;\dot{\b{I}}}$,  we have
\begin{align}
  \b{S''} \circ \dot{\b{E}} + \b{X''} \bullet \dot{\b{I}}
  = \phi_\mathbb{A}(\b{S''},\b{X''}) \pi(\dot{\b{E}}, \dot{\b{I}})
  =
\begin{cases}
\pi  \text{ for } \{ \dot{\b{E}}, \dot{\b{I}} \} \neq \b{0} \\
0 \text{ otherwise}
\end{cases}
   \, .
  \label{polar_functions_2}
\end{align}

From \eqn{legendre_fenchel_2}, the dissipation potential $\pi$ and and the dual dissipation potential $\pi^\star$ are convex conjugates and possess an additive duality, whereas from \eqn{polar_functions_2} the dissipation potential and the canonical yield function $\phi_\mathbb{A}$ have a multiplicative duality.

Thus, in summary, we have
\begin{align*}
  \{ \b{S''} \in \pi_{;\dot{\b{E}}} \, , \,
  \b{X''} \in \pi_{;\dot{\b{I}}} \}
  && \leftrightarrow &&
  \{  \dot{\b{E}}  \in \pi^\star_{;\b{S''}} \, , \,
  \dot{\b{I}} \in \pi^\star_{;\b{X''}} \}
  = N_\mathbb{A}(\b{S''}, \b{X''}) \, ,
\end{align*}
where $ N_\mathbb{A}$ is the normal cone to the admissible region $\mathbb{A}$ as defined in \eqn{normal_cone}.
  \qed
\end{rmk}

\begin{rmk}
  The relation between the non-smooth dissipation potential $\pi$ and the corresponding dual dissipation potential $\pi^\star$ is schematically depicted in \fig{fig_pi_pistar}(a) and (b), for the restricted case where $\pi = \pi(\dot{\b{E}})$ and $\pi^\star = \pi^\star(\b{S''})$.
  The dual dissipation potential $\pi^\star$ is zero for all $\b{S''}\in \mathbb{A}$ and $+\infty$ otherwise (i.e.\ it is the indicator function of the admissible region $\mathbb{A}$).
  The subdifferential of the dissipation potential at the origin contains the set of all possible dissipative stresses interior to the admissible region $\mathbb{A}$ (i.e.\ the region $\interior{\mathbb{A}}$) (see \fig{fig_pi_pistar}(c)).
  Within the admissible region, the magnitude of the dissipative stress is bounded by the yield stress $\sigma_\text{y}$.
  The rate of the generalised strain $\dot{\b{E}}$ lives in the subdifferential to the dual dissipation potential as shown in \fig{fig_pi_pistar}(d).
  \qed
\end{rmk}

\section{Specialisation of the general setting to local plasticity} \label{sec_local_plasticity}

To fix the abstract ideas presented in the previous section on generalised dissipative materials, concrete examples of  energetic and dissipative structures for the familiar problem of local plasticity in three dimensions are presented.
The local plasticity models are constructed by combining simple one-dimensional rheological units.
The resulting models are termed \emph{local serial} and \emph{local parallel} plasticity.
The terminology serial or parallel denotes whether the energetic and dissipative rheological units act in series or in parallel to form the rheological structure.

For the case of local plasticity, the generalised stress $\b{S}$ is the Cauchy stress $\b{\sigma}$, that is  $\b{S} \equiv \{ \b{\sigma} \}$, and the generalised strain $\b{E}$ is the conventional strain $\b{\epsilon}(\b{u}) := \nabla^\sym \b{u}$, that is $\b{E} \equiv \{ \b{\epsilon}(\b{u}) \}$.

\subsection{Local serial plasticity}\label{sec_local_serial_plasticity}

\begin{figure}[h]
  \centering
  \includegraphics[width=\textwidth]{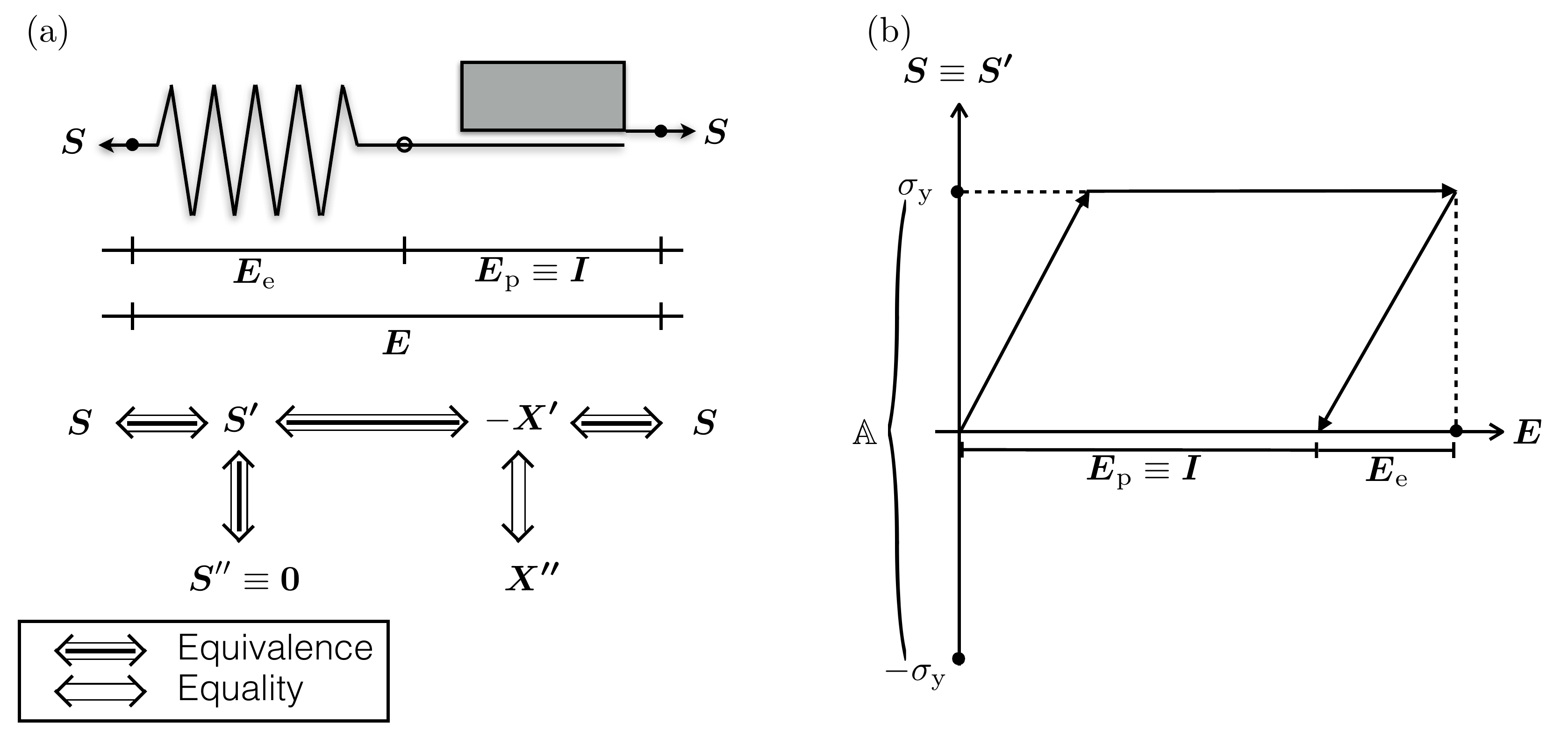}
  \caption{The problem of local serial plasticity. A schematic of the rheological model and the relation between the stress and driving force is shown in (a). In (b), the relation between the stress and the strain is depicted.}
  \label{fig:serial_plasticity}
\end{figure}

Consider the common rheological model of rate-independent \emph{local serial plasticity} presented in \fig{fig:serial_plasticity}(a), where an elastic spring acts in \emph{serial} with a frictional sliding element.
A stress $\b{S}$ is applied to the system and the total strain $\b{E}$ recorded.
The stress-strain response is shown in \fig{fig:serial_plasticity}(b).
A certain magnitude of applied stress (the yield stress $\sigma_\text{y}$) is required to activate the frictional sliding element and thereby induce sliding (plastic flow).
Once the frictional sliding element has yielded, it offerers no further resistance to plastic deformation.
That is, no hardening occurs.
The measure of plastic deformation (sliding) is given by the plastic strain $\b{E}_\text{p}$ which serves as the internal variable.
This rheological model is the prototype for classical rate-independent elasto-perfect-plasticity.

For the case of local serial plasticity, the dissipative stress vanishes, i.e.\ $\b{S}'' \equiv \b{0}$ (see the schematic of the force balance relation in \fig{fig:serial_plasticity}(a)).
The internal variable is associated with the plastic strain, that is $\b{I} \equiv \b{E}_\text{p}$.
The  energy storage potential \eqref{psi} (i.e.\ the energy stored in the elastic spring) is thus given by
\begin{align*}
  \psi = \psi(\b{E},\b{E}_\text{p}) \, .
\end{align*}
A classical \emph{constitutive choice} is to then assume that $\psi = \psi(\b{E} -\b{E}_\text{p}) = \psi(\b{E}_\text{e})$.
Choosing the energy potential to be quadratic in the elastic strain $\b{E}_\text{e}:= \b{E} -  \b{E}_\text{p}$ gives a linear relation between the stress $\b{S}$ and the strain $\b{E}$ in the interior of the admissible region $\interior{\mathbb{A}}:= \{ \b{S} : \phi(\b{S}) < 0 \}$, where the yield function $\phi$ is given here by
\begin{equation}
  \phi(\b{S}) = \varphi( \b{S}) - \sigma_\text{y} 
  =  \vert \b{S} \vert - \sigma_\text{y} \leq 0 \, , \label{yield_local_serial}
\end{equation}
and $\varphi$ is the equivalent stress function. 
Note that in the interior of the admissible  region $\dot{\b{E}}_\text{p} \equiv \b{0}$.
The case of possible plastic flow is given when $\b{S}$ is on the boundary of the admissible region, i.e.\ when $\phi(\b{S})  \equiv 0$.

The total  and energetic stresses, $\b{S}$ and $\b{S'}$ respectively,  are obtained from the elastic law and the generalised stress decomposition  \eqref{S'_from_psi} as
\begin{gather*}
  \b{S'} = \psi_{,\b{E}} \, ,  \\
  \b{S''}\equiv \b{0} = -\b{S}' + \b{S} \quad
  \implies \b{S}' \equiv \b{S} \, .
\end{gather*}
Hence the stress $\b{S}$ and the energetic stress $\b{S'}$ are equivalent for the case of local serial plasticity.

From  definition \eqref{X'_and_X''} of the energetic and dissipative driving forces, $\b{X'}$ and $\b{X''}$ respectively, one obtains the relation between the driving force and the stress as
\begin{gather*}
  \b{X'} = \psi_{,\b{E}_\text{p}}
  \equiv -\psi_{,\b{E}} = - \b{S}' \equiv -\b{S} \, , \\
  \b{X}'' = -\b{X}' \equiv \b{S} \in \pi_{;\dot{\b{E}}_\text{p}}  \, .
\end{gather*}
Hence the dissipative driving force $\b{X''}$ and the stress $\b{S}$ are equivalent for local serial plasticity.
The dissipative driving force can be obtained from the dissipation potential \eqref{diss_S_X_in_pi}.

The reduced dissipation inequality for local serial plasticity follows from \eqn{d_3} as
\begin{align*}
  d = \b{S} \circ \dot{\b{E}}_\text{p} \geq 0 \, .
\end{align*}

It is important to note that the stress $\b{S}$ is fully determined from the elastic law for the model of local serial plasticity.
That is, the energetic force in the spring $\b{S'}\equiv \b{S}$ is known throughout the loading process.

\subsubsection{Prototype non-smooth and smooth dissipation potentials}\label{sec_pi_serial_plasticity}

The prototypical non-smooth dissipation potential \citep[see e.g.][]{Han2010} for local serial plasticity (see \fig{fig_pi_pistar}) is given by
\begin{align}
  \pi(\dot{\b{E}}_\text{p}) = \sigma_\text{y} \vert \dot{\b{E}}_\text{p} \vert \, .
  \label{pi_local_serial}
\end{align}
Note that the non-smooth dissipation potential is a positively homogeneous function of degree 1 (see \eqn{positive_homogeneity}). 
Recall that the choice of a non-smooth dissipation potential corresponds to the rate-independent problem. 

A prototypical regularised form of the dissipation potential\footnote{The results presented in this and subsequent sections hold for any of the widely-used regularisations of the non-smooth dissipation potential.} is given by
\begin{align*}
  \pi(\dot{\b{E}}_\text{p})
  =
  \dfrac{\sigma_\text{y}}{1+\gamma} \vert \dot{\b{E}}_\text{p} \vert^{1+\gamma} \, ,
\end{align*}
where the dissipation parameter $\gamma \in (0,1]$.
The choice of a smooth dissipation potential corresponds to the rate-dependent problem. 
For details of alternative regularisations of the dissipation potential and their associated numerical features, the reader is referred to \citet{Miehe2014b}, among others.
For the limit of $\gamma=0$, the smooth (regularised) potential is equivalent to the non-smooth potential.

\subsection{Local parallel plasticity} \label{sec_local_par_plasticity}
\begin{figure}[h]
  \centering
  \includegraphics[width=\textwidth]{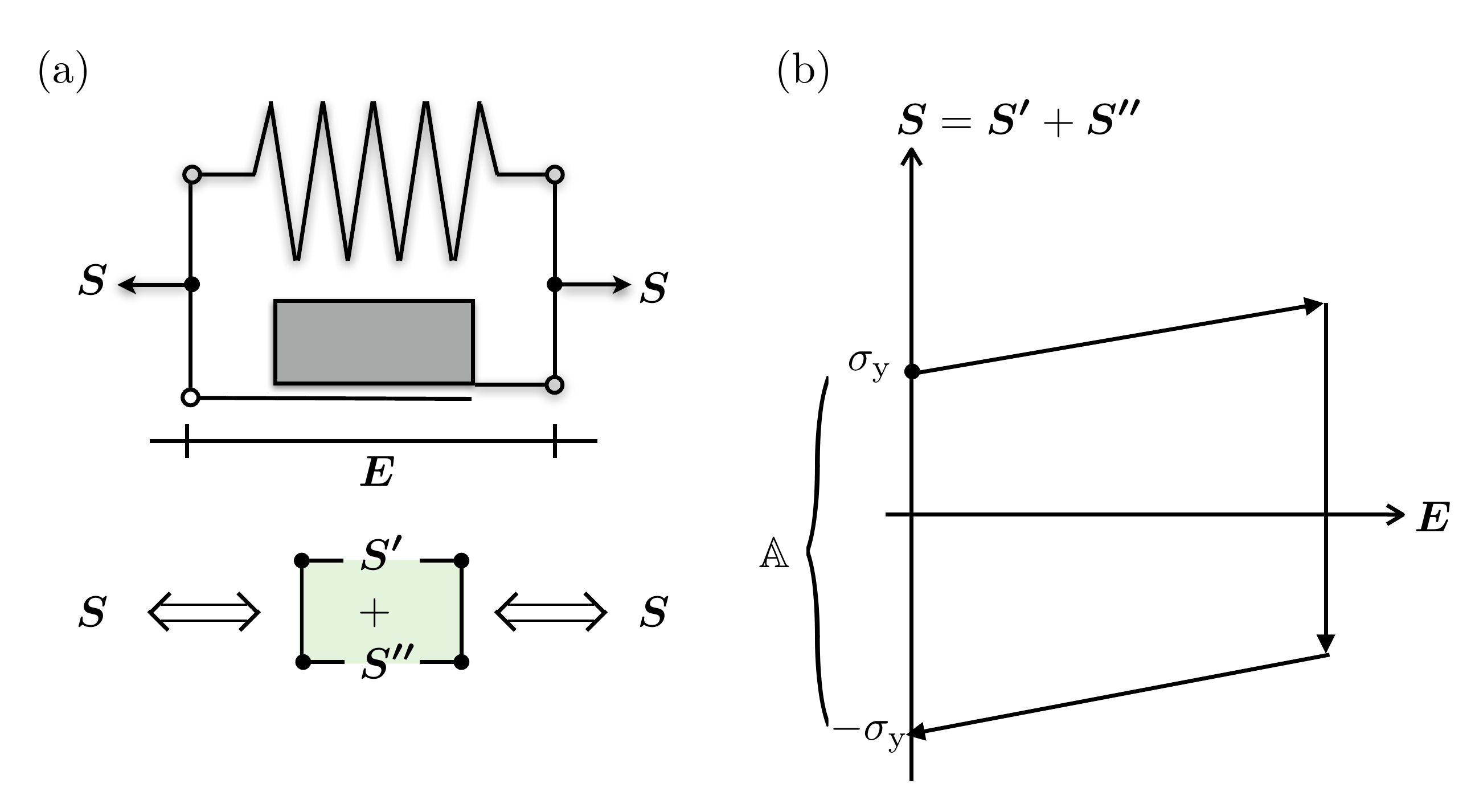}
  \caption{The problem of local parallel plasticity. A schematic of the rheological model and the relation between the energetic and dissipative components of the stress is shown in (a). In (b), the relation between the stress $\b{S}$ and the strain $\b{E}$ is depicted.}
  \label{fig_parallel_plasticity}
\end{figure}

Consider the alternative rheological model of \emph{local parallel plasticity} depicted in \fig{fig_parallel_plasticity}(a).
In local parallel plasticity, an elastic spring and a frictional slider act in parallel to form the rheological unit. 
The model as depicted is rate independent.
The slider will not yield until the magnitude of the applied stress $\b{S}$ reaches the yield stress $\sigma_\text{y}$ and correspondingly  $\phi(\b{S}) = 0$.
Up until the point of yield, the system acts as a rigid body and the split of the stress $\b{S}$ between the slider and the spring can not be determined.
The stress-strain response for the system is shown in \fig{fig_parallel_plasticity}(b).
Upon plastic flow, the spring extends and is responsible for the hardening-type response.
This behaviour corresponds to rigid plasticity. 

For local parallel plasticity, the extension of the spring and of the slider is the same and is given by the total strain $\b{E}$.
The energy storage potential is thus parametrised solely in terms of the strain $\b{E}$, that is
\begin{align*}
  \psi = \psi(\b{E}) \, .
\end{align*}
In addition the dissipative stress $\b{S''} \neq \b{0}$, whereas the set of internal variables is empty, i.e.\ $\b{I} \equiv \b{\emptyset}$.
It is clear from Biot's relation in \subeqn{X'_and_X''}{b} that  $\b{X}'=-\b{X}'' \equiv \b{\emptyset}$.

The reduced dissipation inequality thus follows from \eqn{d_3} as
\begin{align*}
  d = \b{S''} \circ \dot{\b{E}} \geq 0 \, .
\end{align*}

The energetic and dissipative stresses, given by \eqn{S'_from_psi} and \subeqn{diss_S_X_in_pi}{b}, take the form
\begin{align*}
  \b{S'} = \psi_{,\b{E}}
  && \text{and} &&
  \b{S''} \in \pi_{;\dot{\b{E}}}
    && \text{with} &&
  \b{S} = \b{S'} + \b{S''} \, .
\end{align*}
Note, as is clear from the schematic in \fig{fig_parallel_plasticity}(b),  $\b{S''}$ is not determinable in the admissible  region for the model of local parallel plasticity. 
Such a response corresponds to rigid-plastic behaviour.
As will be shown in \sect{sec_micro_parallel} on micromorphic parallel plasticity, the identical problem of an undetermined  generalised stress in the admissible region is present in both the combined energetic-dissipative and the fully-dissipative cases.

\subsubsection{Prototype non-smooth and smooth dissipation potentials}

In a near-identical manner to local serial plasticity (see \sect{sec_pi_serial_plasticity}), the non-smooth dissipation potential for local parallel plasticity is given by
\begin{align*}
  \pi(\dot{\b{E}}) = \sigma_\text{y} \vert \dot{\b{E}} \vert \, .
\end{align*}
Likewise, a prototypical regularised form of the dissipation potential is given by
\begin{align*}
  \pi(\dot{\b{E}})
  =
  \dfrac{\sigma_\text{y}}{1+\gamma} \vert \dot{\b{E}} \vert^{1+\gamma} \, ,
\end{align*}
with the regularisation parameter $\gamma \in (0,1]$.

\section{Micromorphic continua as a framework for strain gradient plasticity} \label{sec_micromorphic_gradient_plasticty}

Micromorphic continua are characterised by additional degrees of freedom associated with each continuum point.
Gradient continua, by contrast, possess higher gradients of the primary fields.
For a classification of a wide range of micromorphic  and gradient plasticity formulations, the reader is referred to \citet{Hirschberger2009, Forest2009, Forest2010}.
The Eringen-type micromorphic framework considered here  can be viewed as a penalised approximation of a Mindlin-type gradient theory (see \ref{sec_mindlin_eringen} for details). 
This choice of framework allows for a unified classification of various important micromorphic and strain gradient plasticity models. 
The additional  \emph{micro} degrees of freedom in the micromorphic theory allow for the modelling of size-dependent phenomena during plastic deformation.

The conventional \emph{macro strain} $\b{\epsilon}(\b{u})$ in a micromorphic continuum is defined, as in the local theory presented in \sect{sec_local_plasticity}, by the symmetric gradient of the displacement field $\b{u}$, that is
\begin{align*}
  \b{\epsilon}(\b{u}) := \nabla^\sym \b{u} \, .
\end{align*}
The \emph{micro strain} and \emph{micro double strain} of the micromorphic continuum are respectively denoted by
\begin{align*}
  \b{\varepsilon} && \text{and} && \b{\gamma}(\b{\varepsilon}):= \nabla \b{\varepsilon} \, .
\end{align*}
To avoid confusion between the similar notation adopted for the macro and micro strain measures, the functional dependence of the macro strain on the displacement will always be indicated.
The two-scale \emph{relative strain} $\b{\delta}$ is defined as the difference between the macro strain $\b{\epsilon}({\b{u}})$ and the micro strain $\b{\varepsilon}$, that is
\begin{align*}
  \b{\delta}({\b{u}}, \b{\varepsilon}) := \b{\epsilon}({\b{u}}) - \b{\varepsilon} \, .
\end{align*}
The magnitude of the relative strain quantifies the closeness of the micromorphic  and the Mindlin-type gradient formulations.
The generalised micro strain $\b{E}$ is defined by
\begin{align*}
  \b{E} := \{ \b{\varepsilon}, \b{\gamma} \ell \} \, ,
\end{align*}
where $\ell > 0$ is an internal length scale.

The internal power density $p^\text{int}_\text{E}$ for the two-field Eringen-type micromorphic continuum considered here is  given by
\begin{align}
  p^\text{int}_\text{E}(\dot{\b{u}}, \dot{\b{\varepsilon}})
  =
  \b{\sigma} : \dot{\b{\delta}}({\b{u}}, \b{\varepsilon})
  + \b{\varsigma}:\dot{\b{\varepsilon}}
  + \b{\mu} \tdot \b{\gamma}(\dot{\b{\varepsilon}}) \, , \label{p_E}
\end{align}
where the conventional \emph{macro stress} is denoted by $\b{\sigma}$, and the \emph{micro stress} and \emph{micro double stress} are denoted by $\b{\varsigma}$ and $\b{\mu}$, respectively.
The generalised micro stress is thus defined by $\b{S}:=\{ \b{\varsigma} , \b{\mu} / \ell \}$.

Following the pioneering work of \citeauthor{Gurtin2002} and co-workers  \citep[see e.g.][]{Gurtin2002, Gurtin2005b}, the corresponding macro and micro governing relations in the absence of body forces  (i.e.\ the Euler--Lagrange equations) are given by
\begin{gather}
  \b{0} = -\div \b{\sigma} \, , \label{macro_force_balance}\\
  \b{\varsigma}  = \b{\sigma} + \div \b{\mu} \, . \label{micro_force_balance}
\end{gather}
The macro relation \eqref{macro_force_balance} is the standard statement of (macro) equilibrium.
The micro relation  \eqref{micro_force_balance} is a two-scale stress balance.\footnote{In order to simplify the presentation, the consequences of the volume preserving nature of plastic flow in metals have not been accounted for here.
In the gradient plasticity theory of \citet{Gurtin2005b}, the macro stress appearing in \eqn{micro_force_balance} is replaced by its deviatoric part.
As a consequence, the micro stress is symmetric and deviatoric, and the micro double stress is symmetric and deviatoric in its first two components.}
The two-scale stress balance is central to the strain gradient plasticity theories of \citeauthor{Gurtin2002} where it is termed the microforce balance.
The microforce balance was derived by \citeauthor{Gurtin2002}, first in the context of single crystal plasticity,  using a virtual power balance.

\citeauthor{Gurtin2002} and co-workers formulate their models solely for the rate-dependent (viscoplastic) case.
The work \citep{Reddy2008} was the first to present the model in a rate-independent framework, using the tools of convex analysis to derive generalisations of the normality condition and, importantly, the notion of a global yield condition.
This was further developed in \citep{Reddy2011_partI, Reddy2011_partII}.
See also \citet{Han2010} for a detailed account.
Much of the material on the choice of a non-smooth dissipation potential presented in the previous sections follows from these works.
A major contribution of the present work is the extension to the smooth case.
In \sect{sec_micro_parallel} the relation between the various strain gradient plasticity theories of \citeauthor{Gurtin2002} and co-workers  and the more general micromorphic setting is made clear.

The resulting dissipation inequality for the two-field Eringen-type micromorphic model reads as follows
\begin{align*}
  d := p^\text{int}_\text{E} - \dot{\psi} \geq 0 \, .
\end{align*}
In the next two sections we analyse various specialisations of the energy storage potential $\psi$ in the micromorphic model.
In the spirit of the local problem presented in \sect{sec_local_plasticity}, these specialisations  are termed \emph{micromorphic serial plasticity} and \emph{micromorphic parallel plasticity}.

\section{Micromorphic serial plasticity}\label{sec_micro_serial}

Serial models for micromorphic plasticity have been presented under different names \citep[see e.g.][]{Sansour2010, Hirschberger2009, Forest2009, Grammenoudis2009a, Grammenoudis2009b}.
For the case of local serial plasticity presented in \sect{sec_local_serial_plasticity}, the energy storage potential was parametrised by the difference between the total strain $\b{E}$ and the plastic strain $\b{E}_\text{p}$.
By analogy, the energy storage potential for micromorphic serial plasticity is given by
\begin{align*}
  \psi =
  \psi( \b{\delta}({\b{u}}, \b{\varepsilon}),
  \lrb{\b{\varepsilon} -  \b{\varepsilon}_\text{p} },
  [\b{\gamma}(\b{\varepsilon}) -  \b{\gamma}_\text{p}] ) \, ,
\end{align*}
where the plastic micro strain and plastic micro double strain are denoted by
\begin{align*}
  \b{\varepsilon}_\text{p} && \text{and} && \b{\gamma}_\text{p} \, .
\end{align*}

The generalised micro stress and plastic micro strain pairs are defined by
\begin{align*}
  \b{S} = \{ \b{\varsigma}, \b{\mu} / \ell \}
  && \text{and} &&
  \b{E}_\text{p} = \{ \b{\varepsilon}_\text{p}, \b{\gamma}_\text{p} \ell \} \ \equiv \b{I} \, ,
\end{align*}
where the internal length scale $\ell > 0$ is introduced, as before, for dimensional consistency.
The generalised inner product of $\b{S}$ and $\dot{\b{E}}_\text{p}$ is thus given by
\begin{align*}
  \b{S} \circ \dot{\b{E}}_\text{p} =
  \b{\varsigma}:\dot{\b{\varepsilon}}_\text{p}
  + \b{\mu} \tdot \dot{\b{\gamma}}_\text{p} \, .
\end{align*}

The macro (energetic) stress $\b{\sigma}$ follows as per the standard definition  as
\begin{align}
  \b{\sigma} = \psi_{,\b{\delta}} \, .
  \label{macro_stress_serial}
\end{align}
As in the local serial model, the generalised micro stress is fully energetic, that is $\b{S} \equiv \b{S'}$ and thus $\b{S''} \equiv \b{\emptyset}$ (see \fig{fig:serial_plasticity}).
The micro stress and micro double stress follow as
\begin{align*}
  \b{\varsigma} = \psi_{,\b{\varepsilon}} = -\psi_{,\b{\varepsilon}_\text{p}} \in \pi_{;\dot{\b{\varepsilon}}_\text{p}}
  && \text{and} &&
  \b{\mu} = \psi_{,\b{\gamma}} = -\psi_{,\b{\gamma}_\text{p}} \in \pi_{;\dot{\b{\gamma}}_\text{p}} \, .
\end{align*}

The generalised micro stress can therefore be expressed as
\begin{align}
  \b{S} = \psi_{,\b{E}} \equiv - \psi_{,\b{E}_\text{p}} \in \pi_{;\dot{\b{E}}_\text{p}} \, ,
  \label{S_micromorphic_serial}
\end{align}
where the subdifferential of the dissipation potential (see the definition \ref{subdifferential})  is given in generalised form as
\begin{align}
  \pi_{;\dot{\b{E}}_\text{p}} :=
  \bigl\{
  \b{S}~\bigr|~\pi(\b{Q}) - \pi(\dot{\b{E}}_\text{p}) - \b{S} \circ [\b{Q} - \dot{\b{E}}_\text{p}] \geq 0 \, , \quad \forall \b{Q}
  \bigr\} \, .\label{sub_diff_micro_plasticity}
\end{align}

The relations characterising micromorphic serial plasticity are identical in structure to those of the familiar local serial plasticity problem presented in \sect{sec_local_serial_plasticity}.
We will now show briefly that the identical conclusions on the determinability of the stress $\b{S}$ in the admissible region for classical local plasticity hold for the generalised stress in the micromorphic case.
The case of a prototypical non-smooth dissipation potential is considered first.

\subsection{Prototype non-smooth dissipation potential}

The prototypical non-smooth dissipation potential for local serial plasticity was given in \eqn{pi_local_serial}.
The corresponding non-smooth dissipation potential (a positively homogeneous function of degree 1) for the problem of micromorphic serial plasticity is given by
\begin{align*}
  \pi(\dot{\b{E}}_\text{p}) = \sigma_\text{y} \vert \dot{\b{E}}_\text{p} \vert \, .
\end{align*}

The  structure of the admissible region $\mathbb{A}$ is a consequence of the choice of the dissipation potential $\pi$.
The interior of the admissible region $\interior{\mathbb{A}}$ is obtained by evaluating the subdifferential of the dissipation potential  \eqref{sub_diff_micro_plasticity} at $\dot{\b{E}}_\text{p} \equiv \b{0}$, which gives
\begin{align*}
  \b{S} \circ \b{Q} \leq \sigma_\text{y} \vert \b{Q} \vert \, , \quad \forall \b{Q} \, .
\end{align*}

The definition of the admissible region of $\b{S}$ then follows from the supremum of the function $r(\b{S};\b{Q})$ given below over all arbitrary $\b{Q}$, that is
\begin{align}
  y(\b{S}) = \sup_{\b{Q}} r(\b{S};\b{Q}) \leq 1
  && \text{where} &&
  r(\b{S},\b{Q}) := \dfrac{\b{S}\circ \b{Q}}{\sigma_\text{y} \vert \b{Q} \vert } \, . \label{local_opt_prob}
\end{align}
The problem of determining the supremum can be stated as a local optimality problem.
First, define $\o{\b{Q}}$ by
\begin{align*}
  \o{\b{Q}} = \arg \biggl\{ \sup_{\b{Q}} r(\b{S};\b{Q}) \biggr\} \, .
\end{align*}
Consequently, we seek the roots of
\begin{align*}
  \dfracp{r(\b{S};\b{Q})}{\b{Q}} \biggr|_{\o{\b{Q}}} \doteq \b{0} \, .
\end{align*}
The local optimality condition follows in the format of an orthogonality condition as
\begin{align}
  \b{S} - r(\b{S};\o{\b{Q}}) \sigma_\text{y} \dfrac{\o{\b{Q}}}{\vert \o{\b{Q}} \vert }
  =
  \b{S} \circ \lrb{\b{I} - \dfrac{\o{\b{Q}}}{\vert \o{\b{Q}} \vert} \otimes \dfrac{\o{\b{Q}}}{\vert \o{\b{Q}} \vert }}
  = \b{0} \, . \label{local_optimality_micro_non_smooth}
\end{align}
The local optimality point is therefore given by $\o{\b{Q}}$ being coaxial to the given $\b{S}$, that is
\begin{align*}
  \dfrac{\b{S}}{\vert \b{S} \vert } \equiv \dfrac{\o{\b{Q}}}{\vert \o{\b{Q}} \vert} \, .
\end{align*}

Thus, inserting $\o{\b{Q}}$ into the function $r(\b{S};\b{Q})$, the admissible region is defined \emph{locally} by
\begin{align}
  y(\b{S}) \equiv \dfrac{\vert \b{S} \vert }{\sigma_\text{y}} \leq 1 \, , \label{local_elas_domain}
\end{align}
which is the canonical yield function $\phi_\mathbb{A} =  y(\b{S})$ of the admissible region $\mathbb{A}$.
The yield function for micromorphic serial plasticity thus has the familiar local structure (c.f.\ \eqn{yield_local_serial}):
\begin{align*}
  \phi(\b{S}) = \varphi(\b{S}) - \sigma_\text{y}  =   \vert \b{S} \vert - \sigma_\text{y} \leq 0 \, .
\end{align*}
The structure of micromorphic serial plasticity is therefore identical to the local problem as depicted in \fig{fig:serial_plasticity}.

For the case of plastic flow (i.e.\ $\dot{\b{E}}_\text{p} \neq \b{0}$), the flow rule follows from the smooth part of the dissipation potential $\pi$ (i.e.\ the region away from the origin) in an identical fashion to local serial plasticity  as
\begin{align*}
  \b{S} = \pi_{,\dot{\b{E}}_\text{p}} =
  \sigma_\text{y} \dfrac{\dot{\b{E}}_\text{p}}{\vert \dot{\b{E}}_\text{p} \vert } \, ,
  \intertext{which can be inverted to obtain}
  \dot{\b{E}}_\text{p} = \dfrac{\vert \dot{\b{E}}_\text{p} \vert \b{S} }{\sigma_\text{y}}
  =:
  \lambda \dfrac{\b{S}}{\vert \b{S} \vert } \, ,
\end{align*}
where $ \varphi(\b{S}) = \vert \b{S} \vert \equiv \sigma_\text{y}$ at yield, and the (positive) plastic multiplier is defined by $\lambda :=  \vert \dot{\b{E}}_\text{p} \vert \geq 0$.
The flow rule for the generalised stress for the case of plastic flow is a positively homogeneous function of degree 0. 
Hence,  $\b{S}(\dot{\b{E}}_\text{p}) =  \b{S}(k \dot{\b{E}}_\text{p})$ for all $k>0$.
The non-smooth dissipation potential therefore corresponds to the rate-independent problem. 

Note that, as for local serial plasticity, $\b{S}$ is fully determined from the elastic law \eqref{S_micromorphic_serial}.

\begin{rmk}\label{rmk_polar}
The definition of the the admissible region for $\b{S}$ in \eqn{local_opt_prob} can be obtained directly from the relation of a polar function given in \ref{sec_polar_functions}.
The support function of $\mathbb{A}$ is given by 
\begin{align*}
\sigma_{\mathbb{A}}(\b{Q}) = \sup_{\b{S}} \{ \b{S} \circ \b{Q} \} \, .
\end{align*}
Then from \eqn{polar_functions}, for $\b{S} \in \mathbb{A}$ and $\b{S} \in \sigma_{\mathbb{A};\b{Q}}=\pi_{;\b{Q}}$, $\b{Q} \neq \b{0}$, we obtain
\begin{align*}
\b{S} \circ \b{Q} &= \phi_{\mathbb{A}}(\b{S}) \sigma_{\mathbb{A}}(\b{Q}) \\
		&= \phi_{\mathbb{A}}(\b{S})  \sigma_\text{y} \vert \b{Q} \vert  \, ,
\end{align*}
where $\phi_{\mathbb{A}}(\b{S})  \equiv y(\b{S}) \leq 1$ is the canonical yield function. 
The expression for the admissible region in \eqn{local_opt_prob} then follows from the supremum of $\b{S} \circ \b{Q} /  \sigma_\text{y} \vert \b{Q} \vert$ over all $\b{Q}$. \qed
\end{rmk}

\subsection{Prototype smooth dissipation potential}

The admissible domain  and the flow law for a prototypical smooth dissipation potential for the case of micromorphic serial plasticity are now determined.
Consider the following regularised dissipation potential (c.f.\ \eqn{pi_local_serial}):
\begin{align*}
  \pi(\dot{\b{E}}_\text{p})
  =
  \dfrac{\sigma_\text{y}}{1+\gamma} \vert \dot{\b{E}}_\text{p} \vert^{1+\gamma} \, ,
\end{align*}
for $\gamma \in (0,1]$.
The admissible domain can be determined by evaluating the subdifferential of the dissipation potential \eqref{sub_diff_micro_plasticity} using the regularised dissipation potential evaluated at $\dot{\b{E}}_\text{p} \equiv \b{0}$.
This gives
\begin{align*}
  \b{S} \circ \b{Q} \leq \dfrac{\sigma_\text{y}}{1+\gamma} \vert \b{Q} \vert^{1+\gamma} \, , \quad \forall \b{Q} \, .
\end{align*}
The admissible domain is obtained by maximising the function $r_\gamma = r_\gamma(\b{S};\b{Q})$ given below over all arbitrary $\b{Q}$ to obtain
\begin{align*}
 y_\gamma(\b{S}) =   
  \max_{\b{Q}}
  \{ r_\gamma(\b{S};\b{Q}) \} \leq\dfrac{1}{1+\gamma}
  && \text{where} &&
  r_\gamma(\b{S};\b{Q}) :=
  \dfrac{\b{S} \circ \b{Q}}{\sigma_\text{y} \vert \b{Q} \vert^{1+\gamma} }
  \, .
\end{align*}
As will be shown in the subsequent remark, the only possible option for the above relation to hold true is for
\begin{align*}
  \b{S} = \b{0} && \implies && \vert \b{S} \vert = 0 \, .
\end{align*}
Thus there is no admissible domain for this choice of smooth dissipation potential, with the consequence that there is always plastic flow, albeit potentially negligible.

\begin{rmk} \label{remark_S_zero_smooth_serial}
  The maximisers $\o{\b{Q}}$ of $r_\gamma(\b{S};\b{Q})$ are given by the following local optimality problem:
  \begin{align}
    \b{S} & \doteq    [1+\gamma ] \lrb{\b{S} \circ\dfrac{\o{\b{Q}}}{\vert \o{\b{Q}} \vert}} \dfrac{\o{\b{Q}}}{\vert \o{\b{Q}} \vert } \, , \label{S_Q_orthogonality}
  \end{align}
  for $\gamma \in (0,1]$.
  For the relation \eqref{S_Q_orthogonality} to hold, it is clear that $\b{S}$ and $\o{\b{Q}}$ must be colinear.
  Thus, the only general choice of $\b{S}$ for $\gamma \in (0,1]$ is $\b{S} = \b{0}$.
  \qed
\end{rmk}

The flow rule then follows directly from the partial derivative of the smooth dissipation potential with respect to $\dot{\b{E}}_\text{p}$ for $\dot{\b{E}}_\text{p} \neq \b{0}$\footnote{For the case of $\dot{\b{E}}_\text{p} = \b{0}$, $\b{S}$ is determined exclusively from the elastic relation or directly using an alternative regularised form of $\pi$ \citep[see e.g.][]{Miehe2014b}.}:
\begin{align*}
  \b{S} = \sigma_\text{y} \vert \dot{\b{E}}_\text{p} \vert^{\gamma} \dfrac{\dot{\b{E}}_\text{p}}{\vert \dot{\b{E}}_\text{p} \vert}
  && \text{with} &&
  \vert \b{S} \vert = \sigma_\text{y} \vert \dot{\b{E}}_\text{p} \vert^{\gamma} \, .
\end{align*}
Hence,
\begin{align*}
  \lrb{\dfrac{\vert \b{S} \vert }{\sigma_\text{y}}}^{1 / \gamma} = \vert \dot{\b{E}}_\text{p} \vert \, .
\end{align*}
Upon rearranging,
\begin{gather*}
  \dot{\b{E}}_\text{p}
  = \lrb{\dfrac{\vert \b{S} \vert}{\sigma_\text{y}}}^{1/\gamma} \dfrac{\b{S}}{\vert \b{S} \vert }
  = \lambda_\gamma \dfrac{\b{S}}{\vert \b{S} \vert }
   \, .
\end{gather*}
Regarding the determination of $\b{S}$, note that from the assumed smoothness of the dissipation potential $\pi$, the flow rule for the generalised stress is an alternative to the elastic law \eqref{S_micromorphic_serial}.

\section{Micrcomorphic parallel plasticity}\label{sec_micro_parallel}

The extension of the local model of parallel plasticity discussed in \sect{sec_local_par_plasticity} to the micromorphic setting is now considered.
As discussed in \sect{sec_micromorphic_gradient_plasticty}, Eringen-type micromorphic plasticity encapsulates the strain gradient plasticity model of \citet{Gurtin2005}.
In the micromorphic model, the total micro strain is the plastic strain in the \citeauthor{Gurtin2005} model, and $\b{\gamma}(\b{\varepsilon})$ is the plastic strain gradient, that is $\b{\varepsilon} \equiv \b{\varepsilon}^\text{p}$ and $\b{\gamma}(\b{\varepsilon}) \equiv \nabla \b{\varepsilon}^\text{p}$.
Thus, the \citeauthor{Gurtin2005} model can viewed as a micrcomorphic parallel plasticity model.

Three specialisations of the micrcomorphic parallel plasticity model are now considered:
\begin{enumerate}[label=(\roman*)]
\item the combined energetic-dissipative case;
\item the hybrid energetic-dissipative case;
\item the fully-dissipative case.
\end{enumerate}
By direct analogy with local  parallel plasticity where the  energy storage potential is parametrised by the total strain $\b{E}$, the inelastic terms in the energy storage potential for the corresponding micromorphic problem relate  to the total micro strain $\b{\varepsilon}$ and its gradient $\b{\gamma}(\b{\varepsilon})$, and the two-scale relative strain $\b{\delta}$.
The energy storage potential is therefore  given by
\begin{align*}
  \psi =
  \psi\bigl( \b{\delta}({\b{u}}, \b{\varepsilon}),
  \b{\varepsilon},
  \b{\gamma}(\b{\varepsilon}) \bigr) \, .
\end{align*}
The dissipation potential is now a function of the rates of the micro strain and micro double strain, that is
\begin{align*}
  \pi = \pi\bigl(\dot{\b{\varepsilon}}, \b{\gamma}(\dot{\b{\varepsilon}})\bigr) \, .
\end{align*}

The macro stress follows as for the serial problem (see \eqn{macro_stress_serial}) as
\begin{align*}
  \b{\sigma} = \psi_{,\b{\delta}} \, .
\end{align*}

The micro stress $\b{\varsigma}$  and micro double stress $\b{\mu}$ are, in general, composed of energetic and dissipative components which can be determined from the energy storage and dissipation potentials as follows:
\begin{align}
  \b{\varsigma} = \b{\varsigma'} + \b{\varsigma''} \in \psi_{,\b{\varepsilon}} + \pi_{;\dot{\b{\varepsilon}}}
  && \text{and} &&
  \b{\mu} = \b{\mu'} + \b{\mu''} \in \psi_{,\b{\gamma}} + \pi_{;\dot{\b{\gamma}}} \, . \label{micro_stress_double_stress_decomp}
\end{align}

The dissipative generalised micro stress $\b{S''} := \b{S} - \b{S'}$ can therefore be defined in terms of the dissipation potential by
\begin{align*}
  \b{S''} &:= \{ \b{\varsigma''}, \b{\mu''} / \ell \} \in \pi_{;\dot{\b{E}}} \, ,
  \intertext{where the generalised micro strain is defined by}
  \b{E} &:= \{ \b{\varepsilon}, \b{\gamma} \ell \} \, .
\end{align*}

The three specialisations of the micrcomorphic parallel plasticity are now presented.
These are obtained by restricting the decomposition of the micro stress $\b{\varsigma}$  and micro double stress  $\b{\mu}$ presented in \eqn{micro_stress_double_stress_decomp}.

\subsection{The combined energetic-dissipative case}

For the combined energetic-dissipative case of micromorphic parallel plasticity, in general, the energetic micro stress $\b{S'} \neq \b{0}$ and, likewise, the dissipative micro stress $\b{S''} \neq \b{0}$.

The admissible region and the flow rule follow from evaluating the subdifferential of the dissipation potential which is given by
\begin{align*}
  \pi_{;\dot{\b{E}}} :=
  \bigl\{
  \b{S''} ~ | ~
  \pi(\b{Q}) - \pi(\dot{\b{E}}) - \b{S''} \circ [\b{Q} - \dot{\b{E}}] \geq 0 \, , \quad \forall \b{Q}
  \bigr\} \, .
\end{align*}
As in the case of local parallel plasticity, in the admissible region the generalised dissipative stress $\b{S''}$ is not determinable from the elastic law for a non-smooth dissipation potential.
The determination of the dissipative stress contribution is similar to the fully-dissipative case discussed  in \sect{sec_fully_diss_micro}, where further details are provided.

\subsection{The hybrid energetic-dissipative case}

Consider now the hybrid energetic-dissipative case where the micro stress $\b{\varsigma} \equiv \b{\varsigma''}$  is fully dissipative and the  micro double stress $\b{\mu} \equiv \b{\mu'}$ is fully energetic, that is:
\begin{align*}
  \b{\varsigma} = \b{\varsigma''} \in \pi_{;\dot{\b{\varepsilon}}}
  && \text{and} &&
  \b{\mu} = \b{\mu'} = \psi_{,\b{\gamma}} \, .
\end{align*}
The hybrid case has been considered by various authors including \citep{Reddy2012}.
For the hybrid case, the microforce balance \eqref{micro_force_balance} becomes
\begin{align*}
  \b{\varsigma''}  &= \b{\sigma} + \div \b{\mu'} \, ,
\end{align*}
and the dissipative micro stress can be determined in the admissible region as both the macro stress and the micro double stress are obtainable  from the energy storage potential.

\subsection{The fully-dissipative case}\label{sec_fully_diss_micro}

For the fully-dissipative problem, the generalised micro-stress $\b{S} \equiv \b{S''}$ and hence
\begin{gather*}
  \b{\varsigma} =  \b{\varsigma''} \in  \pi_{;\dot{\b{\varepsilon}}} \, , \\
  \b{\mu} =  \b{\mu''} \in \pi_{;\dot{\b{\gamma}}} \, , \\
  \b{S} \equiv \b{S''} := \{ \b{\varsigma''}, \b{\mu''} / \ell \} \in \pi_{;\dot{\b{E}}} \, .
\end{gather*}
The fully-dissipative case has been considered by various authors including \citep{Fleck2014, Carstensen2017}.

Once again, the admissible domain and the flow rule follow from evaluating the subdifferential of the dissipation potential, that is
\begin{align*}
  \pi_{;\dot{\b{E}}} :=
  \bigl\{
  \b{S''} ~ | ~
  \pi(\b{Q}) - \pi(\dot{\b{E}}) - \b{S''} \circ [\b{Q} - \dot{\b{E}}] \geq 0 \, , \quad \forall \b{Q}
  \bigr\} \, .
\end{align*}
As was the case for local parallel plasticity and for combined energetic-dissipative micromorphic parallel plasticity, the generalised dissipative stress $\b{S''}$ is not determinable in the admissible region for the fully-dissipative problem with a non-smooth dissipation potential.

The implications of choosing a smooth or a non-smooth dissipation potential on the structure of the admissible region and the flow rule for the fully-dissipative case are now considered.

\subsubsection{Smooth dissipation potential}

Once again, we consider the following canonical regularised dissipation potential
\begin{align*}
  \pi(\dot{\b{E}})
  =
  \dfrac{\sigma_\text{y}}{1+\gamma} \vert \dot{\b{E}} \vert^{1+\gamma} \, ,
\end{align*}
where $\gamma \in (0,1]$.
The admissible region follows from evaluating the subdifferential of the dissipation potential \eqref{sub_diff_micro_plasticity} using the regularised potential $\pi$ at $\dot{\b{E}} \equiv \b{0}$.
This gives
\begin{align*}
  \b{S''} \circ \b{Q} \leq \dfrac{\sigma_\text{y}}{1+\gamma} \vert \b{Q} \vert^{1+\gamma} \, , \quad \forall \b{Q} \, .
\end{align*}
The admissible region is obtained by maximising the  function $r_\gamma = r_\gamma(\b{S};\b{Q})$ over all arbitrary $\b{Q}$ to obtain
\begin{align*}
 y_\gamma (\b{S}) =  \max_{\b{Q}}  r_\gamma(\b{S};\b{Q}) \leq\dfrac{1}{1+\gamma}
  && \text{where} &&
  r_\gamma(\b{S};\b{Q}):= \dfrac{\b{S''} \circ \b{Q}}{\sigma_\text{y} \vert \b{Q} \vert^{1+\gamma} } \, .
\end{align*}
As demonstrated in Remark~\ref{remark_S_zero_smooth_serial} for the smooth  case of micromorphic serial plasticity, the only possible choice of the dissipative micro stress is one where
\begin{gather*}
  \vert \b{S''} \vert = 0 \, .
\end{gather*}
Thus there is no admissible domain for this choice of smooth dissipation potential.

The flow rule then follows directly from the partial derivative of the dissipation potential with respect to $\dot{\b{E}}$ for $\dot{\b{E}} \neq \b{0}$: 
\begin{gather*}
  \b{S''} = \sigma_\text{y} \vert \dot{\b{E}} \vert^{\gamma} \dfrac{\dot{\b{E}}}{\vert \dot{\b{E}} \vert}
  \intertext{or upon rearranging}
  \dot{\b{E}} = \lrb{\dfrac{\vert \b{S''} \vert}{\sigma_\text{y}}}^{1/\gamma} \dfrac{\b{S''}}{\vert \b{S''} \vert } \, .
\end{gather*}
Thus the generalised stress $\b{S} = \b{S''}$ is determinable here from the flow rule when the dissipation potential $\pi$ is smooth.
Furthermore, a smooth dissipation potential allows for a \emph{local} treatment of the flow law and the local (pointwise) determination of the admissible region (which vanishes identically).

\subsubsection{Non-smooth dissipation potential} \label{sec_full_dis_non_smooth_dis_pot}

Consider the prototypical non-smooth dissipation potential
\begin{align}
  \pi(\dot{\b{E}}) = \sigma_\text{y} \vert \dot{\b{E}} \vert \, . \label{full_dis_non_smooth_dis_pot}
\end{align}
The admissible region is obtained by evaluating the subdifferential of the dissipation potential  \eqref{sub_diff_micro_plasticity}, with the non-smooth dissipation potential $\pi$, at $\dot{\b{E}} \equiv \b{0}$.
For this case, the subdifferential of the dissipation potential is given by the set of generalised stresses $\b{S''}$ satisfying 
\begin{align}
  \b{S''} \circ \b{Q} \leq \sigma_\text{y} \vert \b{Q} \vert \, , \quad \forall \b{Q} \, . \label{subdif_dotE_zero}
\end{align}

It follows directly from \eqn{subdif_dotE_zero} that $\b{S''} \leq \sigma_\text{y}$ with $\vert \b{S''}  \vert = \sigma_\text{y}$ at yield.
 The  flow rule  is given by the partial derivative of the dissipation potential with respect to $\dot{\b{E}}$ for $\dot{\b{E}} \neq \b{0}$, that is
\begin{align*}
  \b{S''} = \sigma_\text{y} \dfrac{\dot{\b{E}}}{\vert \dot{\b{E}}\vert }
  \intertext{which can be inverted to obtain}
  \dot{\b{E}} = \dfrac{\vert \dot{\b{E}} \vert \b{S''} }{\sigma_\text{y}}
  =:
  \lambda \dfrac{\b{S''}}{\vert \b{S''} \vert } \, ,
\end{align*}
with the positive plastic multiplier $\lambda \geq 0$.
However, here the generalised stress $\b{S} \equiv \b{S''}$ is not determinable from an elastic law.

Thus, for the fully-dissipative case with a non-smooth dissipation potential, a local definition of the admissible region is not possible.
The implications of the inability to locally determine the admissible region for a non-smooth dissipation potential for the fully-dissipative problem carry over to the combined energetic-dissipative case with a non-smooth dissipation potential.
Given the inability to locally determine the admissible region for theses cases, a possible \emph{global} definition of the admissible domain is thus considered next. 
The global relation was first proposed in \cite{Reddy2008, Reddy2011_partI}.

\subsection{Global dissipation potential for the fully-dissipative problem}

In \sect{sec_micro_parallel} it was shown that for the important case of fully-dissipative micromorphic plasticity with a non-smooth dissipation potential, a local definition of the admissible domain is not possible.
In this section, a global reformulation is presented.
The global relation was first presented in \citep{Reddy2008, Reddy2011_partI}. 
The presentation here differs from that in \citet{Carstensen2017} as we do not resort to a spatial discretisation from the onset.
In order to proceed, we consider first the weak statement of the microforce balance.

\subsubsection{Weak form of the microforce balance}

Recall that for the fully-dissipative case
$\b{\varsigma} =  \b{\varsigma''}$ and  $\b{\mu} =  \b{\mu''}$, that is $\b{S} = \b{S''}$.

The micro boundary conditions considered are the standard ones of either a micro-free Neumann condition $ \b{\mu} \cdot \b{n} \doteq \b{0}$  or a micro-hard Dirichlet condition $ \b{\varepsilon} \doteq \b{0}$ on complementary parts of the boundary $\partial \Omega$ \citep[for more information, see][]{Gurtin2005b}.

The weak form of the microforce balance  is obtained by testing \eqn{micro_force_balance} with an arbitrary  micro strain $\delta \b{\varepsilon}$ which takes zero value on the micro-hard  part of the boundary, to obtain
\begin{align}
  \int_\Omega \b{\sigma}(\b{x}) : \delta \b{\varepsilon}(\b{x}) \, \d x
  =
  \int_\Omega \b{\varsigma}(\b{x}) : \delta \b{\varepsilon}(\b{x}) \, \d x
  -
  \int_\Omega \div \b{\mu}(\b{x}) : \delta \b{\varepsilon}(\b{x}) \, \d x \notag
  \intertext{and by employing  integration by parts}
  \int_\Omega \b{\sigma}(\b{x}) : \delta \b{\varepsilon}(\b{x}) \, \d x
  =
  \int_\Omega \b{\varsigma}(\b{x}) : \delta \b{\varepsilon}(\b{x}) \, \d x
  +
  \int_\Omega  \b{\mu}(\b{x}) \tdot  \b{\gamma}\bigl(\delta \b{\varepsilon}(\b{x})\bigr) \, \d x \, . \label{weak_micro_2}
\end{align}

The generalised macro stress $\b{\Sigma}$ is now defined as a function of position $\b{x}$ by
\begin{align*}
  \b{\Sigma}(\b{x}) := \{ \b{\sigma}(\b{x}),\b{0} \} \, ,
\end{align*}
and an arbitrary generalised micro strain pair by
\begin{align*}
  \delta \b{E}(\b{x})
  =
  \{ \delta \b{\varepsilon}(\b{x}), \b{\gamma}\bigl(\delta \b{\varepsilon}(\b{x})\bigr)  \ell \} \, .
\end{align*}
The weak form of the microforce balance \eqref{weak_micro_2} can thus be expressed as
\begin{align}
  \int_\Omega \b{\Sigma}(\b{x}) \circ \delta \b{E}(\b{x}) \, \d x
  =
  \int_\Omega \b{S}(\b{x}) \circ \delta \b{E}(\b{x}) \, \d x
  \label{weak_micro_force}
\end{align}
Observe that the weak form of the microforce balance allows for the exchange of the generalised micro stress $\b{S}$ with the generalised macro stress $\b{\Sigma}$.
The implications thereof are now presented.

\subsubsection{Exchange of generalised micro and macro stresses}

The global dissipation potential $\Pi$ (a functional) is defined as the integral over the domain  $\Omega$ of the dissipation potential, that is
\begin{align*}
  \Pi \bigl( \dot{\b{E}} \bigr) := \int_\Omega \pi(\dot{\b{E}}(\b{x)}) \, \d x \, .
\end{align*}

The admissible region then follows from evaluating the subdifferential of the global dissipation potential given by
\begin{align*}
  \Pi(\b{Q}) - \Pi(\dot{\b{E}})
  - \int_\Omega \b{S}(\b{x}) \circ [\b{Q}(\b{x}) - \dot{\b{E}}(\b{x})] \, \d x \geq 0 \qquad \forall \b{Q}
  \, .
\end{align*}
The generalized micro stress $\b{S}$ and the generalised macro stress $\b{\Sigma}$ in the subdifferential of the global dissipation potential can be exchanged by virtue of the weak form of the microforce balance \eqref{weak_micro_force}, to give
\begin{align}
  \Pi(\b{Q}) - \Pi(\dot{\b{E}})
  - \int_\Omega \b{\Sigma}(\b{x}) \circ [\b{Q}(\b{x}) - \dot{\b{E}}(\b{x})] \, \d x  \geq 0 \, , \quad \forall \b{Q}
  \, . \label{subdif_glob_dis_pot}
\end{align}

\subsubsection{Non-smooth global dissipation potential}

Recall that a local form of the admissible  region was not possible for the fully-dissipative theory described in \sect{sec_full_dis_non_smooth_dis_pot}.
Consider now the global counterpart to the non-smooth local dissipation potential \eqref{full_dis_non_smooth_dis_pot} given by
\begin{align*}
\Pi(\dot{\b{E}})
  := \int_\Omega \sigma_\text{y}(\b{x}) \vert \dot{\b{E}}(\b{x}) \vert \, \d x \, .
\end{align*}
Note that, in contrast to the presentation in \cite{Carstensen2017}, the yield stress $\sigma_\text{y}$ is  not necessarily assumed uniform.

The structure of the admissible region can be elucidated using the procedure outlined in Remark \ref{rmk_polar}.
By an infinite-dimensional analogue of the result in \ref{sec_polar_functions} \cite{Han2010}, the canonical yield function is given by
\begin{gather}
  Y(\b{\Sigma}) = \sup_{\b{Q}} R(\b{\Sigma};\b{Q}) 
  \leq 1 \qquad \text{where} \qquad
  R(\b{\Sigma};\b{Q}) :=
  \dfrac{\displaystyle \int_\Omega \b{\Sigma}(\b{x}) \circ \b{Q}(\b{x}) \, \d x }{\displaystyle \int_\Omega \sigma_\text{y}(\b{x}) \vert \b{Q}(\b{x}) \vert \, \d x } \, . \label{Y_global}
\end{gather}
Due to the exchange of the generalised micro and macro stresses, the numerator of $R(\b{\Sigma};\b{Q})$ contains the generalised macro stress $\b{\Sigma}$ which is determinable from an elastic law.

The problem of finding the supremum can be expressed as the global optimality problem.
Define the field $\o{\b{Q}}$ by
\begin{align*}
  \o{\b{Q}}
  =
  \arg \biggl\{ \sup_{\b{Q}}
  R(\b{\Sigma};\b{Q})
  \biggr\} \, .
\end{align*}
The global optimality problem follows as
\begin{align*}
  \delta R(\b{\Sigma};\b{Q}) \biggr|_{\o{\b{Q}}} \doteq 0 \, , \quad \forall \b{Q} \, .
\end{align*}
Thus, the global problem for the admissible region results in the variational statement
\begin{align}
  \int_\Omega \b{\Sigma}(\b{x}) \circ \delta \b{Q}(\b{x}) \, \d x
  -
 R(\b{\Sigma};\o{\b{Q}})
  \int_\Omega \sigma_\text{y} \dfrac{\o{\b{Q}}(\b{x})}{\vert \o{\b{Q}}(\b{x}) \vert} \circ \delta \b{Q}(\b{x})  \, \d x = 0  && \forall \delta \b{Q} \, .
  \label{global_elastic_problem}
\end{align}
It is insightful to compare the global problem for the admissible region to the local orthogonality condition \eqref{local_optimality_micro_non_smooth} corresponding to micromorphic serial plasticity.
\eqn{global_elastic_problem} for the admissible domain is global and nonlinear.
 An analytical treatment is not possible.
In principle, a numerical approximation of the problem using the finite element method would be feasible.
Define a trial generalised macro stress  field $ \b{\Sigma}^\star$ as the solution of the macro equilibrium problem \eqref{macro_force_balance} obtained assuming a ``frozen'' plastic state.
\eqn{global_elastic_problem} could then be linearised and solved iteratively for $\o{\b{Q}}$.
An evaluation of the functional \eqref{Y_global} would determine if the trial stress field $\b{\Sigma}^\star$ was indeed admissible.
This form of global algorithm is not conventional and warrants further investigation.

\subsubsection{An upper bound for the global admissible domain}

The numerical approximation of the global admissible region sketched in the previous section is complicated and computationally expensive.
Here we seek a possible estimate for the onset of plastic flow in the form of an upper bound to the global admissible region.

The numerator in the yield functional \eqref{Y_global} can be bounded as follows
\begin{align*}
  \int_\Omega \b{\Sigma}(\b{x}) \circ \b{Q}(\b{x}) \, \d x
  &\leq
  \int_\Omega \dfrac{\vert \b{\Sigma}(\b{x}) \vert}{\sigma_\text{y}(\b{x})} \sigma_\text{y}(\b{x}) \vert \b{Q}(\b{x}) \vert \, \d x \\
  &\leq
  \biggl \Vert \dfrac{\vert \b{\Sigma} \vert}{\sigma_\text{y}} \biggr \Vert_{\infty,\Omega}
  \int_\Omega \sigma_\text{y}(\b{x}) \vert \b{Q}(\b{x}) \vert \, \d x \quad \forall \b{Q} \, .
\end{align*}
Hence,  we define the upper bound estimate for the yield functional by
\begin{align*}
  Y^\star :=
  \biggl \Vert \dfrac{\vert \b{\Sigma} \vert}{\sigma_\text{y}} \biggr \Vert_{\infty,\Omega}
  =
  \biggl \Vert \dfrac{\vert \b{\sigma} \vert}{\sigma_\text{y}} \biggr \Vert_{\infty,\Omega}
  \geq Y(\b{\Sigma}) \, ,
\end{align*}
where $Y(\b{\Sigma}) \leq 1$.
Note that $Y^\star \geq 1$ is possible but does not imply that $Y=1$.
For the case of constant $\sigma_\text{y}$ one obtains $Y^\star \equiv \Vert \vert  \b{\sigma} \vert \Vert_{\infty,\Omega} / \sigma_\text{y}$.

The utility of the upper bound would be as a check for an elastic response.
If $Y^\star < 1$ then the problem is definitely elastic and a numerical approximation for the admissible domain defined in \eqn{global_elastic_problem} need not be sought.

Note that the structure of $Y^\star$ resembles that in \eqn{local_elas_domain} for micromorphic serial plasticity with a non-smooth dissipation potential. 
However, instead of a yield evaluation based on a point-wise absolute value one must determine the maximum over the domain of the ratio of the magnitude of the determinable macro stress to the yield stress.

\section{Conclusion}\label{sec_conclusion}

A unified classification framework for models of extended plasticity has been presented. 
Within this framework, models are classified as either serial or parallel. 
This classification is based on the choice of the energetic and dissipative structures. 
The classification has been introduced first in the familiar setting of local plasticity. 
Prototypical examples of non-smooth and smooth dissipation potentials were examined. 
These correspond to the rate-independent and rate-dependent setting, respectively. 
For non-smooth local parallel plasticity, an identification of the generalised stress prior to the onset of plastic flow is not possible. 
The key results from the local theory were shown to carry over to the extended plasticity models. 

A dissipation-consistent methodology has been adopted throughout to carefully identify the structure of the constitutive  relations and evolution equations. 
In addition, this has ensured that all models are thermodynamically consistent and can be extended to consider other dissipative mechanisms in addition to plasticity. 

The classification has been extended to micromorphic models of plasticity. 
Three categories of micromorphic parallel plasticity have been introduced. 
It has been shown that for the combined energetic-dissipative case and the fully-dissipative case, a local determination of the admissible region for a non-smooth dissipation potential is not possible. 
By contrast, the hybrid energetic-dissipative case permits a local treatment. 
The implications of a global non-smooth dissipation potential for fully-dissipative micromorphic plasticity has been detailed. 

Future work will include the development of algorithms for the fully-dissipative, rate-independent case of micromorphic parallel plasticity. 
An upper bound for the global admissible domain has been proposed.
 However, it was not possible to comment on the sharpness of the bound and hence its utility. 
A careful study of the proposed bound using a smooth dissipation potential for a range of regularisation parameters, including $\gamma \to 0$, will be valuable. 
Such a study will also elucidate the mechanisms that underpin the appearance of an elastic gap \citep[see][]{Fleck2014}. 

The precursor to the proposed numerical investigation is recasting the various models in variational format and then as incremental variational formulations. 
The variational formulation does not have an associated minimisation problem, but the corresponding time-discrete incremental problem does. 
The basis for such an extension is given in \cite{Reddy2011_partI}. 

An extensive classification of available  models of extended plasticity using the proposed scheme  will be the subject of future work. 

\clearpage

\appendix

\section{Standard results and definitions from convex analysis} \label{sec_convex_analysis}

Key results and definitions from convex analysis necessary for the preceding presentation are  summarised here.
For further details refer to \citet{Han2010} and the references therein.

The results are presented in an abstract setting.
$X$ denotes a finite-dimensional normed vector space.
The space of linear continuous functionals on $X$, or dual space of $X$, is denoted by $X^\star$.
For $\b{x} \in X$ and $\b{x}^\star \in X^\star$, the action of $\b{x}^\star$ on $\b{x}$ is defined by the scalar product
\begin{align*}
  \b{x}^\star \cdot \b{x} \, .
\end{align*}
When relating the forthcoming results to the main text it may be useful to make the substitution
\begin{align*}
  \{ \b{S''}, \b{X''} \} \to \b{x}^\star
  && \text{and} &&
  \{ \dot{\b{E}}, \dot{\b{I}} \} \to \b{x} \, .
\end{align*}

\subsection{Convex sets and convex functions}\label{convex_sets_functions}
The subset $Y\subset X$ is \emph{convex}  if
\begin{align}
  \text{for any } \b{x},\b{y} \in Y
  \text{ and }
  0 \leq \theta \leq 1, \quad
  \theta \b{x} + [1-\theta]\b{y} \in Y \, . \label{convex_set}
\end{align}

The \emph{normal cone} to a convex set $Y^\star \subset X^\star$ at $\b{x}^\star$, denoted by $N_{Y^\star}(\b{x}^\star)$, is a \emph{set} in $X$ defined by
\begin{align}
    N_{Y^\star}(\b{x}^\star) :=
    \{
      \b{x} \in X ~:~ \b{x} \cdot [\b{y}^\star - \b{x}^\star] \leq 0
      \, , \quad
      \forall \b{y}^\star \in Y^\star \label{normal_cone}
    \} \, .
\end{align}

The function $f$ is \emph{convex} if
\begin{align}
  f(\theta \b{x} + [1-\theta]\b{y}) \leq \theta f(\b{x}) + [1-\theta]f(\b{y}) \, , \quad
  \forall \b{x},\,\b{y} \in X, \qquad \forall \theta \in [0,1] \, . \label{convex_function}
\end{align}

Given a convex function $f$ on $X$, for any $\b{x} \in X$, the \emph{subdifferential} of $f$ at $\b{x}$, denoted by $f_{;\b{x}}$, is the, possibly empty, subset of $X^\star$ defined by
\begin{align}
  f_{;\b{x}} :=
  \bigl\{ \b{x}^\star \in X^\star \, : \,
  f(\b{y}) \geq f(\b{x}) + \b{x}^\star \cdot [\b{y}-\b{x}] \, , \quad \forall \b{y}\in X \bigr\} \, . \label{subdifferential}
\end{align}
Note that the notation $f_{;\b{x}}$ for the subdifferential used here is not standard. 
A more conventional notation would be $\partial f$.

\subsection{Positive homogeneity}

A function $f(\b{x})$ is said to be  positively homogeneous of degree $k$ if
\begin{align}
  f(\alpha \b{x}) = \vert \alpha^k \vert f(\b{x}) \label{positive_homogeneity}
\end{align}
holds for $k > 0$, where $\alpha \in \mathbb{R}$.

\subsection{Gauge function}

A function $g~:~X \rightarrow [0,\infty]$ is called a gauge if
\begin{subequations} \label{gauge}
  \begin{align}
  & g(\b{x})  \geq 0 \quad \forall \b{x} \in X \, , \label{gauge_i} \\
  & g(\b{0})  = 0 \, , \label{gauge_ii} \\
  & g \text{ is convex and positively homogeneous.} \label{gauge_iii}
\end{align}
\end{subequations}

\subsection{Indicator and support functions}
For $S^\star \subset X^\star$, the \emph{indicator function} $I_{S^\star}$ of $X^\star$ is defined by
\begin{align}
  I_{S^\star}(\b{x}^\star) :=
  \begin{cases}
    0 \, , & \b{x}^\star \in S^\star \, , \\
    +\infty \, , & \b{x}^\star \notin  S^\star \, ,
  \end{cases} \label{indicator_function}
\end{align}
and the \emph{support function} $\sigma_{S^\star}$ is defined on $X$ by
\begin{align}
  \sigma_{S^\star}(\b{x}) := \sup_{\b{x}^\star } \{ \b{x}^\star \cdot \b{x} ~:~ \b{x}^\star \in S^\star \} \, .
  \label{support_function}
\end{align}

For $f$ a function on $X$ with values in $\o{\mathbb{R}} := \mathbb{R} \cup \{ \pm \infty \}$, the \emph{Legendre--Fenchel conjugate} $f^\star$ is the function defined by
\begin{align}
  f^\star(\b{x}^\star) :=
  \sup_{\b{x} \in X}
  \{ \b{x}^\star \cdot \b{x} - f(\b{x}) \} \, , \quad
  \b{x}^\star \in X^\star \, . \label{legendre_fenchel}
\end{align}

Hence the \emph{support function is conjugate to the indicator function}, i.e.\
\begin{align}
  I^\star_{S^\star} = \sigma_{S^\star} 
&& \longleftrightarrow && 
    I_{S^\star} = \sigma^\star_{S^\star} 
  \, . \label{indicator_support_relation}
\end{align}

The following important result relates the support and indicator functions.
Let $K$ be a closed convex set in $X^\star$ defined by
\begin{align}
  K = \{ \b{x}^\star \in X^\star ~:~
  \b{x}^\star \cdot \b{x} \leq g(\b{x}) \} \, ,
\end{align}
where $g$ is a gauge on $X$.
Then
\begin{align}
  \begin{split}
  g(\b{x}) &= \sigma_K(\b{x}) \, , \\
  g^\star(\b{x}^\star) &= I_K(\b{x}^\star) \, , \\
  K &= g_{;\b{x}}(\b{0}) \, , \\
  \b{x}^\star & \in g_{;\b{x}}
  \quad \longleftrightarrow \quad
  \b{x}  \in g^\star_{;\b{x}^\star} = N_K(\b{x}^\star) \, .
\end{split}
\end{align}

\subsection{Polar functions} \label{sec_polar_functions}

Let $K \subset X^\star$ be a closed convex set whose boundary, denoted  $\bdy{K}$, is the level set $c_0$ of the convex function $\varphi(\b{x}^\star)$.
That is
\begin{align}
  K = \{ \b{x}^\star \in X^\star ~:~
  \varphi(\b{x}^\star) \leq c_0 \} \, ,
\end{align}
for $c_0 > 0$.
The function $\varphi$ can be defined so that it is a gauge on the set $K$, denoted by $\phi_K$, where
\begin{align*}
	K = \bigl\{ \b{x}^\star ~:~ \phi_K(\b{x}^\star) \leq 1 \bigr\} \, .
\end{align*}
It can be shown that for $\b{x}^\star \in K$ and $\b{x}^\star \in \sigma_{K;\b{x}}$, $\b{x} \neq \b{0}$ we have
\begin{align}
  \b{x}^\star \cdot \b{x} = \phi_K(\b{x}^\star) \sigma_K(\b{x}) \, . \label{polar_functions}
\end{align}
Hence $\phi_K$ and $\sigma_K$ are polar conjugates of each other.

\section{Mindlin-, Hu--Washizu- and Eringen-type theories}\label{sec_mindlin_eringen}

\subsection{Mindlin-type gradient continuum}

The internal power density of a Mindlin-type gradient continuum \citep{Mindlin1964} is given by
\begin{align}
p^\text{int}_\text{M}(\dot{\b{u}}) &:=
\b{\varsigma} : \gradsym \dot{\b{u}}
+ \b{\mu} \tdot  \grad \lrrb{ \gradsym \dot{\b{u}}}  \, , \label{p_Mindlin_i}
\end{align}
where $\b{u}$ is the displacement,  $\b{\varsigma}$ is the stress tensor, and $\b{\mu}$ is the double stress tensor.
The corresponding single-field Euler--Lagrange equation in the absence of body forces is given by
\begin{align}
-\div \lrrb{\b{\varsigma} - \div \b{\mu}} = \b{0} \, . \label{EL_Mindlin}
\end{align}
The Euler--Lagrange equation is a fourth-order partial differential equation in terms of the displacement field $\b{u}$.
The strain tensor $\b{\epsilon}(\b{u})$  is defined   in terms of the symmetric displacement gradient by
\begin{align*}
\b{\epsilon}(\b{u}) := \gradsym \b{u} \, .
\end{align*}

\subsection{Three-Field Hu--Washizu formulation}

A three-field Hu–-Washizu \citep{Hu1955, Washizu1982} type-formulation for a gradient continuum is obtained by introducing an independent symmetric tensor field $\o{\b{\varepsilon}}$ constrained as follows
\begin{align*}
\o{\b{\varepsilon}} \doteq \gradsym{\b{u}} \, ,
\intertext{and whose gradient is denoted by}
\o{\b{\gamma}} := \grad \o{\b{\varepsilon}} \, .
\end{align*}
The equality of $\o{\b{\varepsilon}}$ and $\gradsym{\b{u}}$ can be expressed in the form of a constraint by
\begin{align}
\o{\b{\delta}}(\b{u},\o{\b{\varepsilon}}) := \gradsym{\b{u}}  - \o{\b{\varepsilon}}  \doteq \b{0} \, . \label{HW_constraint}
\end{align}
The constrained variables are distinguished by an overbar.

The internal power density corresponding to a three-field Hu--Washizu formulation for a gradient continuum is given by
\begin{align}
p^\text{int}_\text{HW}(\dot{\b{u}},\dot{\o{\b{\varepsilon}}}, \b{\lambda}) :=
\b{\varsigma}:\dot{\o{\b{\varepsilon}}} + \b{\mu}\tdot\dot{\o{\b{\gamma}}} + \b{\lambda}:\dot{\o{\b{\delta}} } \, , \label{P_int_HW}
\end{align}
where $\b{\lambda}$ is the tensorial Lagrange multiplier to impose the constraint (\ref{HW_constraint}).

The Euler--Lagrange equations corresponding to \eqn{P_int_HW} follow as
\begin{subequations}
\begin{align}
-\div \b{\lambda} &= \b{0} \, , \label{EL_HW_a} \\
\div \b{\mu} &= \b{\varsigma} - \b{\lambda} \, . \label{EL_HW_b}
\end{align}
\end{subequations}
The Euler--Lagrange equations corresponding  to a Mindlin-type gradient continuum  (\ref{EL_Mindlin}) are recovered by  substituting the expression for the Lagrange multiplier in \eqn{EL_HW_b} into \eqn{EL_HW_a}.
It is thus clear that the solution to the three-field formulation given by (\ref{EL_HW_a} -- \ref{EL_HW_b}) is equivalent to that of the one-field formulation (\ref{EL_Mindlin}).

\subsection{Eringen-type micromorphic continua}

The strict enforcement of the constraint (\ref{HW_constraint}) via the Lagrange multiplier $\b{\lambda}$ can be relaxed via a penalisation of the constraint violation whereby
\begin{align*}
\b{\lambda} \to \b{\sigma} = \b{\sigma}({\b{\delta} })
&& \text{with} &&
{\b{\delta} } := \gradsym{\b{u}}  - {\b{\varepsilon}}  \neq \b{0} \, .
\end{align*}

As a consequence of relaxing the constraint, the  internal power density becomes that of a two-field Eringen-type  formulation, given by
\begin{align}
p^\text{int}_\text{E}(\dot{\b{u}},\dot{{\b{\varepsilon}}}) :=
{\b{\varsigma}}:\dot{{{\b{\varepsilon}}}} + {\b{\mu}}\tdot\dot{{{\b{\gamma}}}} + \b{\sigma}:\dot{{\b{\delta} } } \, . \label{P_int_E}
\end{align}
The corresponding Euler--Lagrange equations describing an Eringen-type micromorphic continuum are given by
\begin{subequations}
\begin{align}
-\div \b{\sigma} = \b{0} \, , \label{EL_E_a} \\
\div {\b{\mu}} = {\b{\varsigma}} - \b{\sigma} \label{EL_E_b} \, .
\end{align}
\end{subequations}
The solution to the Eringen-type micromorphic formulation (\ref{EL_E_a}--\ref{EL_E_b}) only coincides asymptotically with that of the Mindlin-type gradient formulation as $ {\b{\delta}} \to  {\b{0}}$.
%

\section*{Acknowledgements}

AM and PS thanks the Engineering and Physical Sciences Research Council for their support through the  Strategic Support Package: Engineering of Active Materials by Multiscale/Multiphysics Computational Mechanics (grant reference number 300129).

  BDR acknowledges the support of the National Research Foundation  of South Africa through the South Africa Research Chair in Computational Mechanics.
  
  PS thanks the German Research Foundation (DFG) for funding his research through the Collaborative Research Centre 814 and the Priority Programme 2013.

\clearpage
\section*{References}
\bibliographystyle{unsrtnat}
\bibliography{bib_file}

\end{document}

%% file: packages_style_defs.tex
\usepackage{graphicx,psfrag,epsfig,rotating}
\usepackage{color}

\usepackage{booktabs}
\usepackage{url}
\usepackage{longtable}

\usepackage{multirow}
\usepackage{setspace}
\usepackage{nomencl}
\usepackage{lscape}
\usepackage{tabularx}
\usepackage{pdflscape}

\usepackage{array,float,latexsym,amsmath,amssymb,amsbsy,amsthm,eurosym,theorem,amsfonts,mathrsfs,bbm,verbatim}

\usepackage[active]{srcltx}

\usepackage{pdfsync}

\newcommand{\dfracp}[2]{\dfrac{\partial #1}{\partial#2}}

\newcommand{\eqn}[1]{Eq.~(\ref{#1})}

\newcommand{\subeqn}[2]{Eq.~(\ref{#1})${}_\text{#2}$}
\newcommand{\fig}[1]{Fig.~\ref{#1}}

\newcommand{\sect}[1]{Sec.~\ref{#1}}

\renewcommand{\b}[1]{\boldsymbol{#1}} 

\renewcommand{\o}[1]{\overline{#1}}

\newcommand{\lrb}[1]{\left[ {#1} \right]}
\newcommand{\lrrb}[1]{\left( {#1} \right)}

\newcommand{\bdy}[1]{\mathop{\mathrm{bdy}({#1})}}
\newcommand{\interior}[1]{\mathop{\mathrm{int}({#1})}}

\newcommand{\sym}{{\mbox{\scriptsize sym}}}

\renewcommand{\d}{\mbox{d}}

\newcommand{\grad}{\nabla}

\renewcommand{\div}{\mbox{div}}

\newcommand{\gradsym}{\nabla^\sym }

\newcommand{\trns}{{}^{\mathsf{\scriptsize T}}}

\definecolor{gray}{rgb}{0.75, 0.75, 0.75}
\definecolor{yellow}{rgb}{1, 0.7, 0.2}
\definecolor{green}{rgb}{0.3, 0.9, 0.3}
\definecolor{brown}{rgb}{0.6, 0.3, 0.2}
\definecolor{magenta}{rgb}{0.9, 0.1, 0.9}
\definecolor{light}{rgb}{1, 0.7, 0.7}

